\documentclass[10pt]{article}
\usepackage{amssymb}
\usepackage[a4paper, total={6in, 8.3in}]{geometry}
\usepackage{sidecap}
\usepackage{float}
\usepackage{multirow}
\usepackage[utf8]{inputenc}
\usepackage{latexsym}
\usepackage{enumerate}
\usepackage{caption}
\usepackage{wrapfig}
\usepackage[perpage]{footmisc}
\usepackage{subcaption}
\usepackage[hyperref]{xcolor}
\usepackage{epsfig}
\graphicspath{{projectplots/}}
\renewcommand\thefootnote{\textcolor{nicered}{\fnsymbol{footnote}}}

\usepackage{relsize}
\usepackage{indentfirst}
\usepackage{amssymb}
\usepackage{bookmark}
\usepackage{amsthm}
\usepackage{amsmath}

\makeatletter
\renewcommand\@dotsep{300}   
\makeatother

\usepackage{rotating}

\usepackage{tabularx} 
\usepackage{amsmath}  
\usepackage{graphicx} 
\usepackage{cite} 
\usepackage{hyperref} 

\definecolor{azure(colorwheel)}{rgb}{0.0, 0.5, 1.0}
\definecolor{deepmagenta}{rgb}{0.8, 0.0, 0.8}
\definecolor{cerulean}{rgb}{0.0, 0.48, 0.65}
\definecolor{nicered}{rgb}{0.7,0.1,0.1}
\definecolor{darkpastelgreen}{rgb}{0.01, 0.75, 0.24}
\definecolor{nicegreen}{rgb}{0.1,0.5,0.1}
\definecolor{MyDarkBlue}{rgb}{0,0.1,0.7}
\definecolor{DarkBlue}{RGB}{0,0,153}
\definecolor{PetuniaColor}{RGB}{150,23,147}
\definecolor{myblue}{RGB}{10,10,200}
\definecolor{Purple}{RGB}{128,0,128}

\hypersetup{colorlinks,bookmarksopen,bookmarksnumbered,
citecolor={nicered},
linkcolor={MyDarkBlue},pdfstartview=FitH,
urlcolor={cerulean}}

\usepackage{titlesec}
\titlespacing*{\section}
{0pt}{1\baselineskip}{0.8\baselineskip}

\usepackage{scalerel}
\usepackage{tikz}
\usetikzlibrary{svg.path}
\definecolor{orcidlogocol}{HTML}{A6CE39}
\tikzset{
  orcidlogo/.pic={
    \fill[orcidlogocol] svg{M256,128c0,70.7-57.3,128-128,128C57.3,256,0,198.7,0,128C0,57.3,57.3,0,128,0C198.7,0,256,57.3,256,128z};
    \fill[white] svg{M86.3,186.2H70.9V79.1h15.4v48.4V186.2z}
                 svg{M108.9,79.1h41.6c39.6,0,57,28.3,57,53.6c0,27.5-21.5,53.6-56.8,53.6h-41.8V79.1z M124.3,172.4h24.5c34.9,0,42.9-26.5,42.9-39.7c0-21.5-13.7-39.7-43.7-39.7h-23.7V172.4z}
                 svg{M88.7,56.8c0,5.5-4.5,10.1-10.1,10.1c-5.6,0-10.1-4.6-10.1-10.1c0-5.6,4.5-10.1,10.1-10.1C84.2,46.7,88.7,51.3,88.7,56.8z};}}
\newcommand\orcid[1]{\href{https://orcid.org/#1}{\mbox{\scalerel*{
\begin{tikzpicture}[yscale=-1,transform shape]
\pic{orcidlogo};
\end{tikzpicture}
}{|}}}}

\def\beq{\begin{equation}}
\def\eeq{\end{equation}}
\def\bea{\begin{eqnarray}}
\def\eea{\end{eqnarray}}

\newcommand{\myref}[2]{{\color{myblue}\ref{#1}(\subref{#2})}}

\newcommand{\newc}{\newcommand}
\def\eq$#1${\begin{equation}#1\end{equation}}
\def\gat$#1${\begin{gather}#1\end{gather}}
\def\bal$#1${\begin{align}#1\end{align}}
\def\eqarr$#1${\begin{eqnarray}#1\end{eqnarray}}
\newc{\pa}{\partial}
\newc{\alp}{\alpha}
\newc{\gam}{\gamma}
\newc{\Gam}{\Gamma}
\newc{\del}{\delta}
\newc{\eps}{\epsilon}
\newc{\lam}{\lambda}
\newc{\sig}{\sigma}
\newc{\ups}{\upsilon}
\newc{\ome}{\omega}
\newc{\pphi}{\varphi}
\newc{\nonum}{\nonumber}
\newc{\hami}{\text{\textbf{\lat{H}}}}
\newc{\gren}{\mathcal{G}}
\newc{\lagr}{\mathcal{L}}
\newc{\timor}{\mathcal{T}}
\newc{\prop}{\mathcal{K}}
\newc{\zcal}{\mathcal{Z}}
\newc{\cinf}{\mathcal{C}_\infty}
\newc{\operx}{\text{\textbf{\lat{x}}}}
\newc{\opera}{\text{\textbf{\lat{a}}}}
\newc{\operp}{\text{\textbf{\lat{p}}}}
\newc{\operl}{\text{\textbf{\lat{L}}}}
\newc{\gfv}{g^{(5)}}
\newc{\kfv}{\kappa_{(5)}}
\newc{\tf}{\tilde{f}}
\newc{\tlam}{\tilde{\Lambda}}
\newc{\tl}{\tilde{\lam}}
\newc{\dist}{\displaystyle}
\newc{\ra}{\rightarrow}
\newc{\Ra}{\Rightarrow}
\newc{\hsp}{\hspace{1em}}
\newc{\wtild}{\widetilde}
\newc{\ssst}{\scriptscriptstyle}

\newc{\pan}[1]{\textcolor{deepmagenta}{#1}}
\newc{\teo}[1]{\textcolor{azure(colorwheel)}{#1}}
\newc{\corr}[1]{\textcolor{red}{#1}}

\begin{document}

\begin{titlepage}

\vspace*{3cm}

\begin{center}

{\bf \LARGE Analytic and exponentially localized brane-world Reissner-Nordstr\"{o}m-AdS solution: a top-down approach}

\bigskip \bigskip \medskip

{\bf Theodoros Nakas\orcid{0000-0002-3522-5803}}\footnote[1]{\,t.nakas@uoi.gr} and
{\bf Panagiota Kanti\orcid{0000-0002-3018-5558}}\footnote[2]{\,pkanti@uoi.gr} 

\bigskip
{\it Division of Theoretical Physics, Department of Physics,\\
University of Ioannina, Ioannina GR-45110, Greece}

\bigskip \medskip
{\bf Abstract}
\end{center}
In this work, we construct a five-dimensional spherically-symmetric, charged and asymptotically
Anti-de Sitter black hole with its singularity being  point-like and strictly localised on our brane.
In addition, the induced brane geometry is described by a Reissner-Nordstr\"{o}m-(A)dS line-element.
We perform a careful classification of the horizons, and demonstrate that all of them are exponentially 
localised close to the brane thus exhibiting a pancake shape.
The bulk gravitational background is everywhere regular, and reduces to an AdS$_5$ spacetime
right outside the black-hole event horizon. This geometry is supported by an anisotropic fluid
with only two independent components, the energy density $\rho_E$ and tangential pressure
$p_2$. All energy conditions are respected close to and on our brane, but a local violation takes place 
within the event horizon regime in the bulk. A tensor-vector-scalar field-theory model is built in an
attempt to realise the necessary bulk matter, however, in order to do so,  both gauge and scalar
degrees of freedom need to turn phantom-like at the bulk boundary. The study of the junction
conditions reveals that no additional matter needs to be introduced on the brane for its consistent
embedding in the bulk geometry apart from its constant, positive tension. We finally compute
the effective gravitational equations on the brane, and demonstrate that the
Reissner-Nordstr\"{o}m-(A)dS geometry on our brane is caused by the combined effect of the
five-dimensional geometry and bulk matter with its charge being in fact a tidal charge.

\end{titlepage}

\renewcommand{\thefootnote}{\arabic{footnote}}

\setcounter{page}{1}

\numberwithin{equation}{section}

\vspace*{0em}
\hspace{-1em}\rule{17.3cm}{0.1mm}
\vspace{-5em}
\hypersetup{linktocpage}
\tableofcontents
\hspace{-1em}\rule{17.3cm}{0.1mm}

\vspace*{1em}

\section{Prologue}

The idea of the existence of extra spacelike dimensions, first suggested by Kaluza \cite{Kaluza} and
Klein \cite{Klein}, is now more than a hundred years old. During this time interval, it has been employed 
in order to formulate fundamental theories of particle physics, such as string theory \cite{Green, Polchinski}, or more
phenomenologically-oriented models such as the Large Extra Dimensions \cite{ADD1, AADD, ADD2}
or Warped Extra Dimensions \cite{RS1, RS2} scenaria. In the latter two models in particular, our four-dimensional
world is a 3-brane \cite{misha, akama} embedded in a higher-dimensional spacetime (the bulk). 
In the Large Extra Dimension scenario   \cite{ADD1, AADD, ADD2}, the extra spacelike dimensions
are compactified to a new length scale---this scale is an independent  scale of the theory
which however has to be smaller than the $\mu m$ scale in order to avoid observation 
\cite{Kapner, Franc}. In the Warped Extra Dimensions scenario\cite{RS1, RS2}, there is only one extra spacelike
dimension, which may be either compactified, by introducing a second brane, or assumed to be infinite---in 
the latter case, the localisation of the graviton close to our brane, ensured by an exponentially decaying
warp factor in the metric, leads to an effective compactification of the infinite fifth dimension.

Motivated by the above, a plethora of higher-dimensional studies in both particle physics and gravity
were performed over the years. The interest in the construction  of black-hole solutions 
living in an arbitrary number of spacelike dimensions, flat or warped, was intense. In the case of Large
Extra Dimensions, the assumed flatness of the higher-dimensional spacetime allowed for previously
derived, analytical black-hole solutions \cite{Tangherlini, MP} to accurately model such
gravitational objects also in the new context. In contrast, the warping of the bulk spacetime in the
case of the Warped Extra Dimensions scenario posed a significant obstacle in the construction of
analytical solutions describing black holes centered on our brane and extending in a regular bulk
spacetime. In the first such attempt \cite{CHR}, the effort to construct a five-dimensional 
brane-world black hole led instead to the emergence of a black-string solution; although the
line-element on the brane matched the Schwarzschild solution, its singularity was not point-like
and localised on our brane, where the gravitational collapse had taken place, but it was extending
along the infinite fifth dimension. This infinitely-long black-string solution was plagued by a
curvature singularity at the AdS infinity, thus refuting the gravity localisation, and was also proven
to be unstable \cite{GL, Ruth}. Since then, numerous attempts for the construction of a robust
brane-world black-hole solution have been made in the literature (for a partial only list, see
\cite{EHM, Dadhich, tidal, Papanto, KT, Dadhich2, CasadioNew, EFK, Frolov, EGK, Tanaka, KOT,
Charmousis, Kofinas, Shanka, Karasik, GGI, CGKM, Ovalle, Harko, daRocha1, AS, Fitzpatrick, Zegers,
Heydari, Dai, Bruni,  Yoshino, Kleihaus, daRocha2, Andrianov1, Andrianov2, Cuadros, KPZ, KPP,
Banerjee, Chakraborty1,Chakraborty2,Chakraborty3}). 
Despite these efforts, an exact, analytic solution describing a  five-dimensional black hole 
localised close to our brane and leading also to a Schwarzschild black hole on the brane was never
found. The construction, on the other hand, of numerical solutions describing small \cite{Kudoh1, Kudoh2}
and large \cite{Tanahashi, Figueras1, Page}  black holes as well as the easiness with which black-string
solutions emerge in the context of the same theory (again, for a non-exhaustive list, see
\cite{Gubser, Wiseman, Kudoh3, Sorkin1, Sorkin2, Kleihaus2, Headrick, Charmousis1, Charmousis2,
Figueras2, Kalisch, Emparan2, Cisterna1, KNP1, Cisterna2, Cisterna3, CFLO, Rezvanjou,
KNP2, KNP3, Estrada, Cisterna2021}) kept vibrant the interest in providing an answer as to what
type of geometrical construction  could lead to the long-sought brane-world black-hole solution. 

In a recent work of ours \cite{NK1}, we have addressed the above question and constructed from first principles,
the geometry of an analytic, spherically-symmetric five-dimensional black hole. This was done by
combining both bulk and brane perspectives, that is by employing a set of coordinates that ensured 
the isotropy of the five-dimensional spacetime and combining it with an appropriately selected metric
function of the four-dimensional line-element. This geometrical construction resulted into a black-hole
solution which had its singularity strictly localised at a single point on the brane. Its horizon was extending
into the bulk, as expected, but it had a pancake shape and was localised exponentially close to our brane.
The five-dimensional background was everywhere regular and reduced to a pure AdS$_5$ spacetime 
right outside the black-hole horizon. This geometric solution was not a vacuum one, and a form of bulk
matter had to be introduced in order to support it. However, this bulk matter was characterised by a 
very simple energy-momentum tensor describing an anisotropic fluid with only two independent components,
the energy density and tangential pressure. The geometric background on the brane was of a pure
Schwarzschild form, which was shown to satisfy the gravitational field equations of the effective 
four-dimensional theory. 

In the present work, we generalise our previous analysis by retaining the basic method for the construction
of the five-dimensional, spherically-symmetric black hole but by considering an alternative form of the 
metric function. This form is inspired by the one of the four-dimensional Reissner-Nordstr\"{o}m-(A)dS 
solution. In this way, we allow for the presence of a charge term and of a cosmological constant in the
effective metric, thus generalising our previous assumption of a neutral, asymptotically-flat brane
black hole. However, being also part of a five-dimensional line-element, the richer topological structure
following from this new metric function is transferred also in the bulk. Thus, we perform a thorough
study of both the horizon structure of the five-dimensional spacetime and of all curvature invariants. 
We demonstrate that the singularity of the black hole remains point-like and strictly localised on the
brane. We also show that every horizon radius characterising the spacetime, depending on the values
of its parameters, acquires a pancake shape and gets exponentially localised close to the brane. 
The five-dimensional spacetime is everywhere regular and reduces again to an  AdS$_5$ spacetime 
right outside the black hole event horizon. 
Our analysis remains at all points analytical and manages to accommodate all geometrical features
necessary for the localisation of a physical black-hole solution close to our brane. In fact, as we will
demonstrate, our constructed line-element has the same general structure as the one used in 
\cite{Figueras1} for the numerical construction of such a solution  with the difference that in our
case all metric components are analytically known.

We next turn to the question of what is the form of the bulk matter that would support the
aforementioned  geometry. To this end, we perform a detailed study of the bulk energy-momentum
tensor, and show that its minimal structure with only two independent components is preserved also
in this case. We then attempt to construct a field-theory model for
the realisation of the bulk matter, in the form of a five-dimensional tensor-vector-scalar theory, 
and discuss the conditions under which such a description could be viable. We then focus on the
presence of the brane itself, and we first study the junctions conditions which govern its consistent
embedding  in the five-dimensional background. We demonstrate that these are satisfied by a
brane with no additional matter apart from its positive tension. Finally, we derive the 
gravitational equations of the effective theory and demonstrate that they are indeed satisfied
by the induced solution on the brane, namely the Reissner-Nordstr\"{o}m-(A)dS solution.

The structure of this paper is as follows: in Section \ref{sec: GS}, we present the general method for
constructing the five-dimensional geometry and study its geometrical properties. In
Section \ref{sec: GT}, we turn to the gravitational theory, study the profile of the bulk matter and 
present the field-theory toy model. In Section \ref{sec: JC-ET}, we investigate the junction conditions
and the effective gravitational theory on the brane. We summarise our analysis and
discuss our results in Section \ref{sec: Ep}. 

\section{The Geometrical Setup}
\label{sec: GS}

We start our analysis with the Randall-Sundrum (RS) metric ansatz  \cite{RS1, RS2} which describes
a five-dimensional warped spacetime. Its line-element has the form 
\eq$\label{rs-metric1}
ds^2=e^{-2k|y|}\left(-dt^2+d\vec x ^{\,2}\right)+dy^2\,.$
The aforementioned spacetime is comprised by four-dimensional flat slices
stacked together along a fifth dimension denoted by the coordinate $y$. The warp
factor of each slice is $e^{-2k|y|}$, where $k$ is the curvature of the five-dimensional
AdS spacetime. In  \cite{RS1, RS2}, the AdS bulk spacetime is sourced by a negative
five-dimensional cosmological constant.  In the context of this work, we will
assume that our four-dimensional world is represented by a 3-brane located at
$y=0$. 

We may also write the above line-element in conformally-flat coordinates: we thus
introduce a new bulk coordinate $z$ via the relation $z = sgn(y)\,(e^{k |y|}-1)/k$.
In addition, as we will address the presence of localised, spherically-symmetric solutions
on our brane, we employ spherical coordinates for the spatial brane directions. Then, the
line-element \eqref{rs-metric1} takes the form
\eq$\label{rs-metr-z}
ds^2=\frac{1}{(k|z|+1)^2}\left(-dt^2+dr^2+r^2\,d\Omega_2^2+dz^2\right)\,,$
where $d\Omega_2^2=d\theta^2+ \sin^2\theta\,d\phi^2$. As in the original RS models, the bulk-related 
${\mathbf Z}_2$ symmetry under the change $z \rightarrow -z$ has been also preserved here for
consistency reasons.

\begin{figure}[t!]
\centering
\includegraphics[width=0.7 \textwidth]{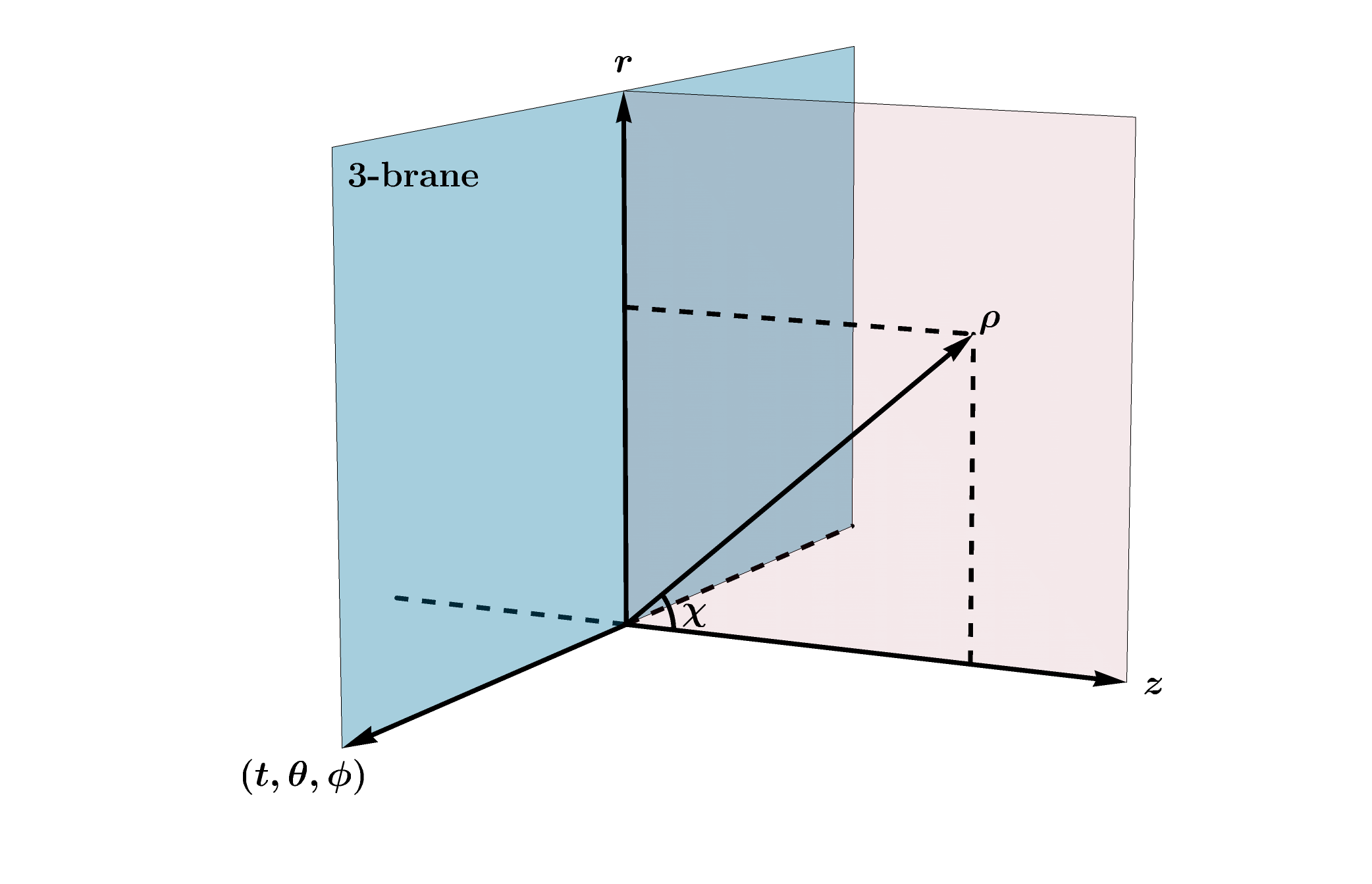}
\vspace{-2em} 
\caption{The geometrical set-up of the five-dimensional spacetime and the set of coordinates.}
\vspace{-0.5em}   
\label{coords-plot}
\end{figure}

We will now extend the spherical symmetry into the bulk by performing the following change of variables
\eq$\label{new-coords}
\left\{ r=\rho\sin\chi\,,\hspace{0.5em}z=\rho\cos\chi\right\}\,,$
in terms of which  Eq. \eqref{rs-metr-z} reads
\eq$\label{rs-metr-sph}
ds^2=\frac{1}{(1+k\rho|\cos\chi|)^2}\left(-dt^2+d\rho^2+\rho^2\,d\Omega_3^2\right)\,.$
In the above, $\chi$ is an angular coordinate which takes values in the range $[0, \pi]$, $\rho$ denotes the radial 
coordinate of the four-dimensional spatial part of the five-dimensional spacetime, while 
\eq$d\Omega_3^2=d\chi^2+\sin^2\chi\,d\theta^2+\sin^2\chi\,\sin^2\theta\,d\phi^2$ 
corresponds to the line-element of a unit 3-sphere.
From \eqref{new-coords} we easily deduce that $\rho$ is always positive definite while
the domains $\chi\in\left[0,\pi/2\right)$ and $\chi\in\left(\pi/2,\pi\right]$ correspond to positive
and negative values of $z$, respectively. The line-element \eqref{rs-metr-sph} is invariant
under the coordinate transformation $\chi \rightarrow \pi - \chi$, which relates the two domains,
due to the assumed ${\mathbf Z}_2$ symmetry. We may therefore focus only on the $\left[0,\pi/2\right]$
domain for which  $\cos\chi \geq 0$. The inverse transformation to \eqref{new-coords} reads
\eq$\label{new-coords-inv}
\big\{\rho= \sqrt{r^2+z^2}\ , \quad
\tan \chi=r/z\big\}\,.$
From this, we deduce that the radial coordinate $\rho$ ranges over the interval $[0, \infty)$. 
In fact, it  receives contributions both from the (brane) $r$ and (bulk) $z$ coordinates. Therefore,
the radial infinity, $\rho \rightarrow \infty$, may describe both the asymptotic
AdS boundary ($|z| \rightarrow \infty$) and the radial infinity on the brane ($r \rightarrow \infty$). 
In the new coordinate system \eqref{new-coords}, the brane, which was located at $y=0$ or $z=0$,
now lies at $\cos \chi=0$, i.e. at $\chi =\pi/2$. There, $\rho$ reduces to the brane radial coordinate $r$. 
The geometrical set-up of the five-dimensional spacetime along with the two systems of
coordinates are depicted in Fig. \ref{coords-plot}. 
 
As mentioned previously, we are interested in placing a spherically-symmetric black hole
on our brane. To this end, we replace the two-dimensional flat part $(-dt^2 +d\rho^2)$
of the line-element in Eq. \eqref{rs-metr-sph} with a curved part, and thus we write
\eq$\label{5d-metr}
ds^2=\frac{1}{(1+k\rho \cos\chi)^2}\bigg[-f(\rho)\,dt^2+\frac{d\rho^2}{f(\rho)}+
\rho^2\,d\Omega_3^2\bigg]\,, \hspace{1.5em}\chi\in[0,\pi/2]\,.$
Here, $f(\rho)$ is a general spherically-symmetric function. In the context of this work,
we are interested in the study of black holes, and we will therefore assume that $f(\rho)$
has a form inspired by the more general spherically-symmetric black-hole solution of General
Relativity, namely the Reissner-Nordstr\"{o}m-(Anti-)de Sitter solution:
\eq$\label{ansatz}
f(\rho)=1-\frac{2M}{\rho}+\frac{Q^2}{\rho^2}-\frac{\Lambda}{3}\, \rho^2\,.$
Note that, on the brane where $\cos\chi=0$ and $\rho=r$, the line-element \eqref{5d-metr} 
{\it does} reduce to a Reissner-Nordstr\"{o}m-(Anti-)de Sitter black hole, with the parameter $M$
being related to its mass, $Q$ to its charge and $\Lambda$ to the effective cosmological
constant on the brane.\footnote[1]{A similar construction of the bulk geometry was followed in \cite{Dai},
however, a different form was used for the function $f(\rho)$. As a result, no known black-hole
solution was recovered on the brane. In addition, their choice did not support either an AdS$_5$
spacetime asymptotically in the bulk, in contrast with our choice as we will shortly demonstrate.}

\begin{figure}[t!]
    \centering
    \begin{subfigure}[b]{0.49\textwidth}
        \includegraphics[width=\textwidth]{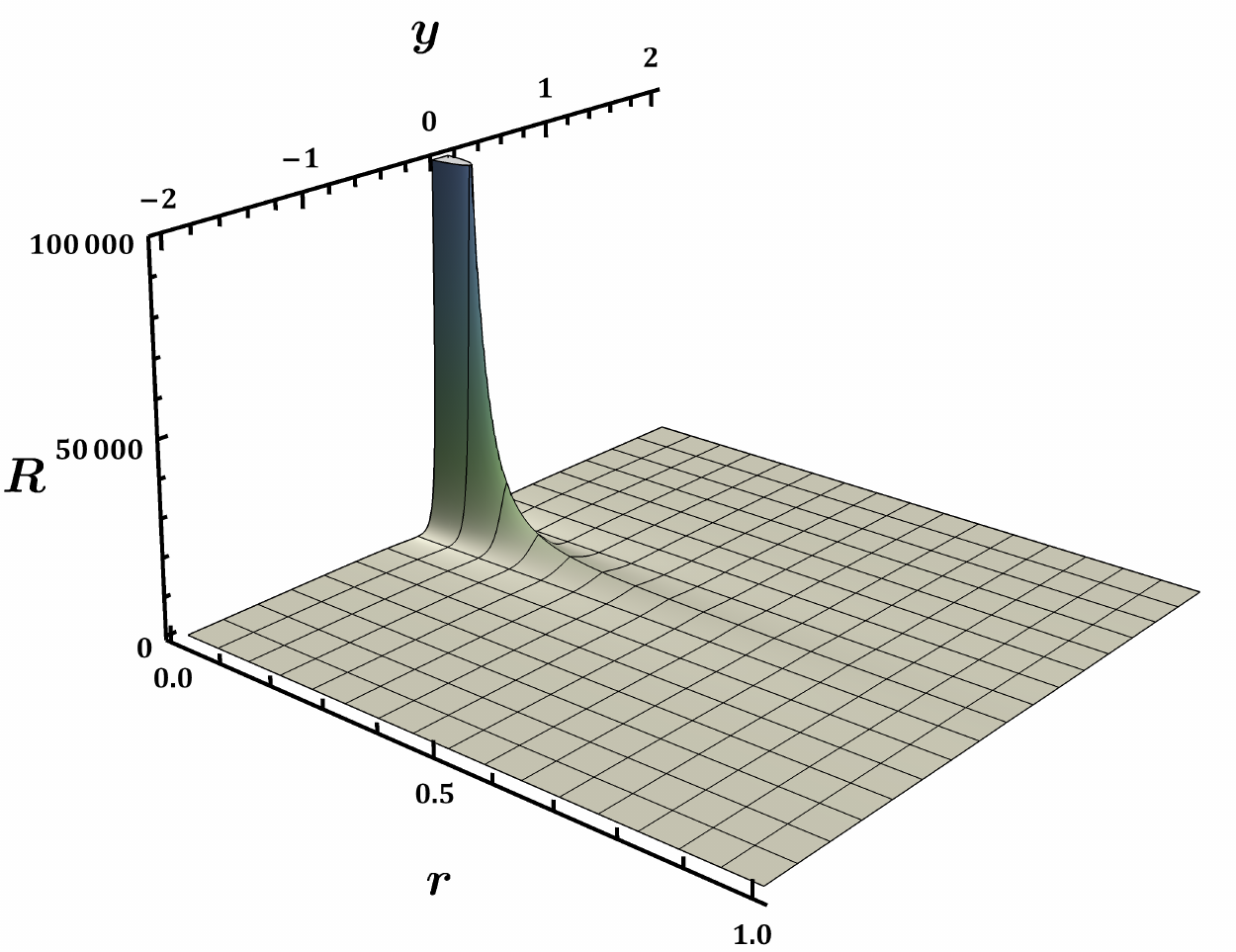}
        \caption{\hspace*{0em}}
        \label{Ricci1-1}
    \end{subfigure}
    \hfill
    ~ 
    \begin{subfigure}[b]{0.48\textwidth}
        \includegraphics[width=\textwidth]{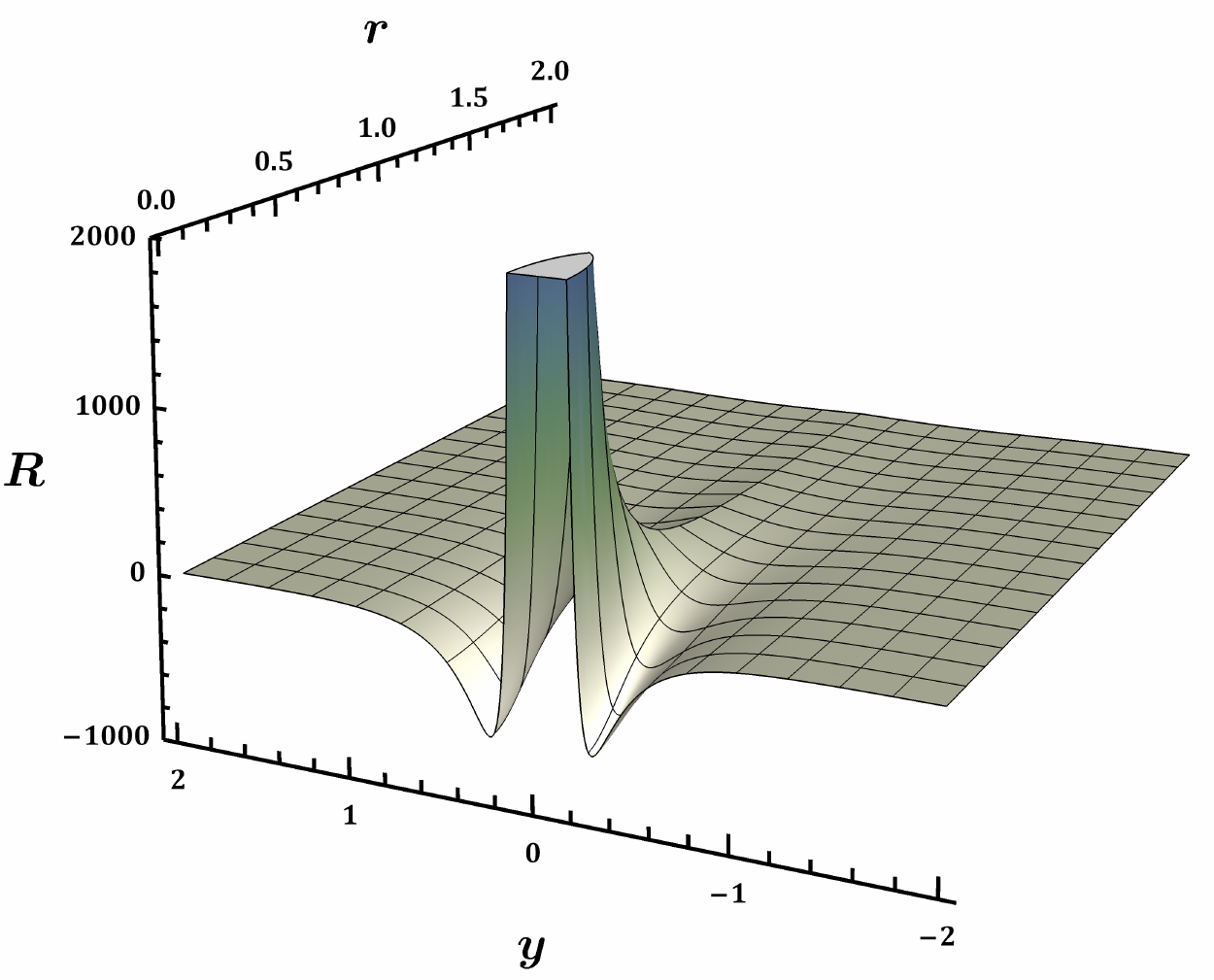}
        \caption{\hspace*{2.9em}}
        \label{Ricci1-2}
    \end{subfigure}  
    \vspace{-0.5em}  
    \caption{(a) The scalar curvature $R$ in terms of the coordinates $(r,y)$ for 
    $k=1$, $M=10$, $Q=1$, and $\Lambda=5\times 10^{-4}$, while (b) shows a magnification of the geometry near the singularity.    }
    \label{Ricci1-plot}
\end{figure}

However, its interpretation from the bulk point-of-view needs to be carefully examined. 
Indeed, almost all known analytic black-hole solutions on the brane either lack completely a bulk description,
or correspond to bulk solutions with an undesired topology (i.e. that of a black string) or unattractive characteristics
(i.e. non-asymptotically AdS solutions). We will therefore investigate now the topological characteristics
of our five-dimensional construction. To this end, we compute all scalar gravitational
quantities, namely the Ricci scalar $R$, the Ricci tensor combination $\mathcal{R} \equiv R^{MN}R_{MN}$ 
and the Kretchmann scalar $\mathcal{K} \equiv R^{MNKL}R_{MNKL}$. The expression of the Ricci
scalar is the most elegant one and is given below
\eq$\label{ricci-1}
R=-20 \left(k^2-\frac{\Lambda}{3} \right)+\frac{12k^2M\cos^2 \chi}{\rho }-\frac{12k\cos\chi \left(k Q^2 
\cos \chi+2 M\right)}{\rho ^2}+\frac{4\left(2 k Q^2 \cos \chi+ M\right)}{\rho ^3}\,,$
while the more extended  $\mathcal{R}$ and $\mathcal{K}$ quantities are presented in Appendix \ref{app: Curv-Inv}. 
The above expression contains a constant term which involves the warping parameter $k$ and the effective
cosmological parameter $\Lambda$. It also contains additional terms sourced by the mass and charge
of the black hole. These terms are singular at the value $\rho=0$
of the bulk radial coordinate. However, this singularity arises only when $r$ and $z$ are {\it simultaneously}
zero, i.e. at the location of the black-hole singularity {\it on the brane}. Any bulk point having by definition a non
zero value of $z$, and thus a non-zero value of $\rho$, is regular. In addition, all singular terms vanish in the
limit $\rho \rightarrow \infty$, i.e. 
when approaching the AdS asymptotic boundary or the radial infinity on the brane. Therefore, the spacetime
\eqref{5d-metr} does describe the gravitational background around a five-dimensional localised black hole with
a spacetime singularity strictly restricted on the brane. We also note that no singularity arises at the AdS
asymptotic boundary, a feature which plagues most non-homogeneous  black-string solutions. In our case,
far away from the brane, the spacetime becomes a maximally-symmetric one with a curvature determined
by the combination $-20\,(k^2-\Lambda/3)$. For $\Lambda=0$, we obtain the exact same  AdS spacetime 
of the Randall-Sundrum model; for positive but small values---compared to the warping effect driven by 
$k$---of the effective cosmological constant on the brane, the AdS character of the asymptotic regime is again
retained\,\footnote[2]{Although mathematically possible, we do not consider here the case where $\Lambda > 3k^2$.
Since $k$ is an energy scale close to the fundamental gravity scale, that would demand an extremely large
$\Lambda$. Such an assumption is not supported by current observational data.} while, for $\Lambda<0$,
it is further enhanced.

\begin{figure}[t!]
    \centering
    \begin{subfigure}[b]{0.49\textwidth}
        \includegraphics[width=\textwidth]{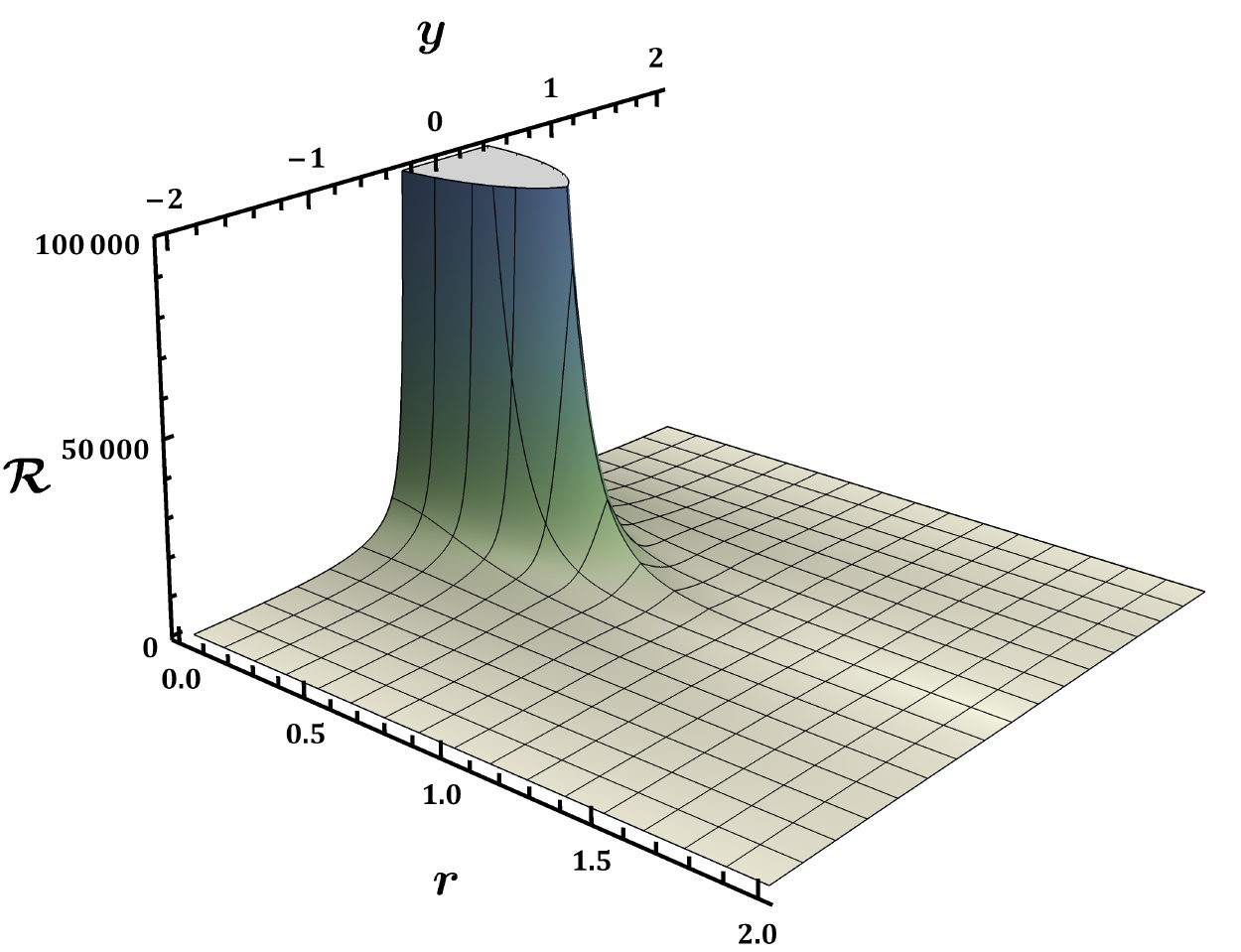}
        \caption{\hspace*{0em}}
        \label{Ricci2}
    \end{subfigure}
    \hfill
    ~ 
    \begin{subfigure}[b]{0.48\textwidth}
        \includegraphics[width=\textwidth]{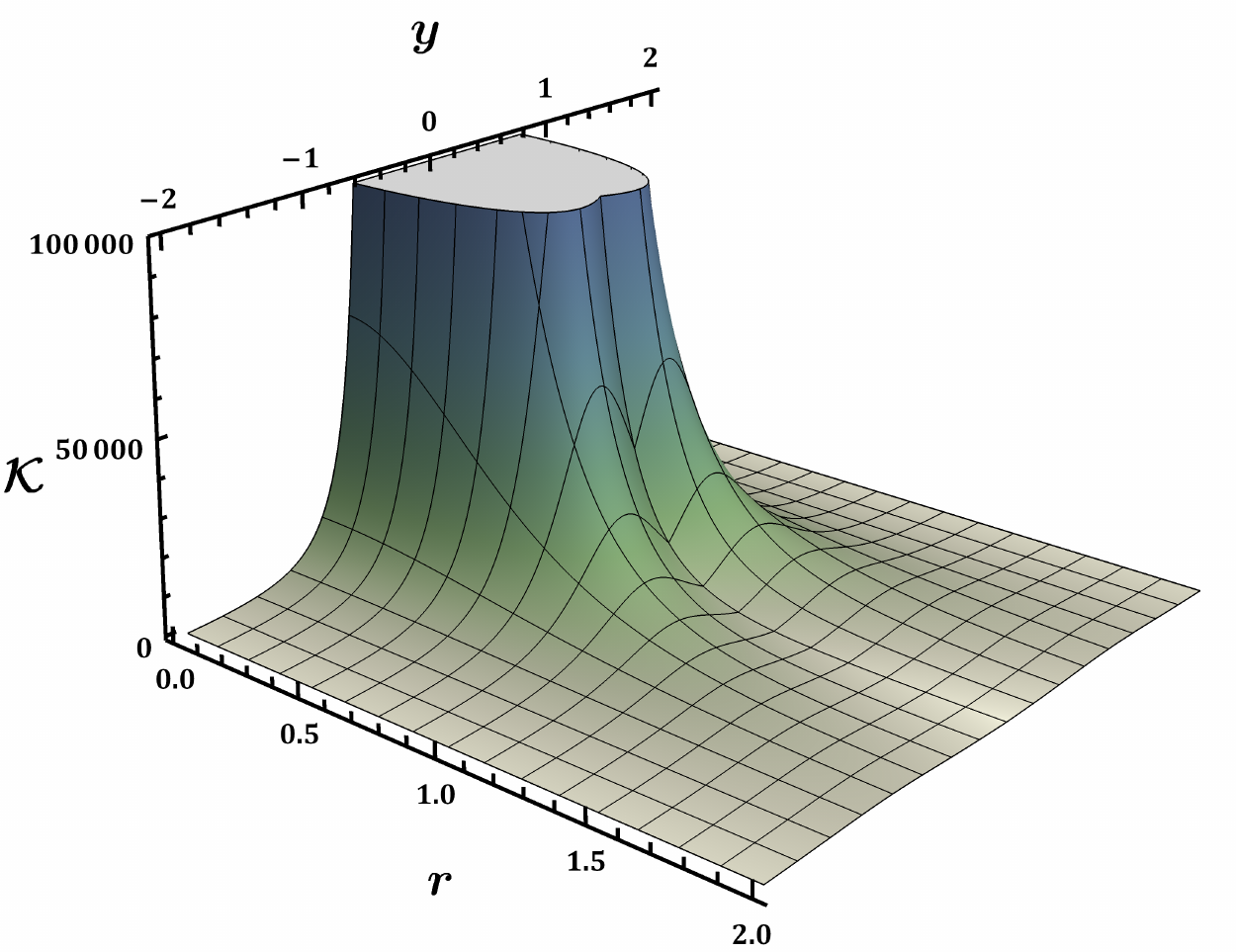}
        \caption{\hspace*{0em}}
        \label{RiemSq}
    \end{subfigure}  
    \vspace{-0.5em}  
    \caption{(a) The invariant quantity $\mathcal{R}\equiv R_{MN}R^{MN}$ in terms of the coordinates $(r,y)$ for 
    $k=1$, $M=10$, $Q=1$, and $\Lambda=5\times 10^{-4}$, and (b) the invariant quantity $\mathcal{K}\equiv R_{MNKL}R^{MNKL}$
    for the same values of the parameters.    }
    \label{Ricci2-Riem-plot}
\end{figure}

The expressions of the $\mathcal{R}$ and $\mathcal{K}$ invariant quantities  displayed in Appendix \ref{app: Curv-Inv}
also lead to the same conclusions drawn above for the topology of the five-dimensional spacetime.
It is illuminating to plot the behaviour of all curvature quantities. To this end, we use the original
$(r,y)$ brane and bulk coordinates as it is easier to depict the location of the brane. Using \eqref{new-coords-inv}
in \eqref{ricci-1}, we easily obtain for $R$ the expression
\eq$\label{ricci-1-ry}
R=-20 \left(k^2-\frac{\Lambda}{3} \right)+\frac{4k^3M\left(10-12e^{k|y|}+3e^{2k|y|}\right)}{\left[k^2r^2
+\left(e^{k|y|}-1\right)^2\right]^{3/2}}-\frac{4k^4Q^2\left(5-8e^{k|y|}+3e^{2k|y|}\right)}{\left[k^2r^2
+\left(e^{k|y|}-1\right)^2\right]^2}\,.$
Similar expressions may be derived for $\mathcal{R}$ and $\mathcal{K}$, and these are again presented in
Appendix \ref{app: Curv-Inv}. In Fig. \ref{Ricci1-plot}, we depict the Ricci scalar $R$ in terms of both $r$ and 
$y$---we remind the reader that, in this coordinate system, the brane is located at $y=0$. We observe that the
curvature of the 5-dimensional spacetime increases {\it only} when we approach the brane {\it and simultaneously}
take the limit $r \rightarrow 0$. All other bulk or brane points are regular. The curvature quickly decreases
as we move away from the singularity on the brane acquiring its constant, negative, asymptotic value 
corresponding to an AdS spacetime---this value is much smaller than the one adopted in the
vicinity of the singularity and thus is not visible in the plots. In  Fig. \myref{Ricci1-plot}{Ricci1-2}, we present a
magnification of the behaviour of the Ricci scalar close to the singular point: we observe the presence
of an interesting regime in the bulk where the curvature of spacetime dips to a large negative value before 
starting to increase close to the singularity. We will comment on this feature in the following
section. In Figs. \myref{Ricci2-Riem-plot}{Ricci2} and \myref{Ricci2-Riem-plot}{RiemSq}, we also present the 
behaviour of the $\mathcal{R}$ and $\mathcal{K}$ invariant quantities, respectively. They exhibit the same
asymptotic and near-singularity behaviours as the scalar curvature $R$ with the only difference being the
absence of the negative curvature well. 

In order to discuss further the topology of the five-dimensional spacetime \eqref{5d-metr}, let us also re-write 
it in terms of the original non-spherical coordinates $(r,y)$. Employing again the inverse transformations
(\ref{new-coords-inv}), the line-element takes the form
\bal$\label{metr-r-y}
ds^2=e^{-2k|y|}&\left\{-f(r,y) dt^2+\frac{dr^2}{r^2+z^2(y)}\biggl[\frac{r^2}{f(r,y)}+z^2(y)\biggr]
+r^2d\Omega_2^2+\frac{2r z(y)\,e^{k|y|}}{r^2+z^2(y)}\biggl[\frac{1}{f(r,y)}-1\biggr]drdy \right\}\nonum\\[1mm]
&\hspace{0.7em}+\frac{dy^2}{r^2+z^2(y)}\left[r^2+\frac{z^2(y)}{f(r,y)}\right]\,,$
where $z(y)=sgn(y)(e^{k |y|}-1)/k$, and 
\eq$\label{f-ry}
f(r,y)=1-\frac{2M}{\sqrt{r^2+z^2(y)}}+\frac{Q^2}{r^2+z^2(y)}-\frac{\Lambda}{3}\big[r^2+z^2(y)\big]\,.$
As a side remark, we note that  the line-element \eqref{metr-r-y} has come out to have the {\it exact same}
structure as the one employed in \cite{Figueras1} for the numerical construction of a 5-dimensional black
hole localised on the brane. In our case though all metric components are analytically known whereas, in
\cite{Figueras1}, the five unknown metric fuctions had to be numerically determined.

Returning to the topology of our five-dimensional solution, we are interested in the behaviour of the
black-hole horizon(s) in the bulk. If the aforementioned spacetime
describes a regular, localised-on-the-brane black hole, its horizon(s) are expected to extend into the bulk but
stay close to the brane. To investigate this, we will study the causal structure of the bulk spacetime as this
is defined by the light cone. We will consider radial null trajectories in the five-dimensional background
\eqref{metr-r-y}, and thus keep $\theta$ and $\phi$ constant. Then, for a fixed value $y=y_0$ of the fifth
coordinate, the condition $ds^2=0$ leads to the result
\eq$ \label{null-traj}
\frac{dt}{dr}=\pm\frac{1}{f(r,y_0)}\left[\frac{r^2k^2+f(r,y_0)\left(e^{k|y_0|}-1\right)^2}{r^2k^2+
\left(e^{k|y_0|}-1\right)^2}\right]^{1/2},$
where
\eq$\label{horizon-function}
f(r,y_0)=1-2M\left[r^2+\frac{\left(e^{k|y_0|}-1\right)^2}{k^2}\right]^{-\frac{1}{2}}+Q^2\left[r^2
+\frac{\left(e^{k|y_0|}-1\right)^2}{k^2}\right]^{-1}-\frac{\Lambda}{3}\left[r^2+\frac{\left(e^{k|y_0|}-1\right)^2}{k^2}\right].$
The location and topology of the horizons characterising the line-element \eqref{metr-r-y} may be obtained
via Eq.  \eqref{null-traj}, by determining the values of $(r,y_0)$ for which $dt/dr=\pm\infty$, or equivalently
$f(r,y_0)=0$. For $y_0=0$, Eq. \eqref{horizon-function} reduces to the metric function $f(r)$ of a four-dimensional
Reissner-Nordstr\"{o}m-(anti-)de Sitter spacetime for which the emergence and location of horizons has been
extensively studied (see, for example, \cite{Chambers,Bousso}). A similar analysis may be conducted
also in the context of the five-dimensional spacetime  \eqref{metr-r-y}, where the location of horizons is
determined by the equation $f(\rho)=0$, with the bulk radial coordinate being $\rho = \sqrt{r^2 + z^2(y)}$ ---we 
keep the $y$-coordinate fixed in Eqs. \eqref{null-traj} and \eqref{horizon-function} in order to present
the view of a static ``observer'' located at different slices of the bulk spacetime as we move away from the brane. 

\begin{sidewaysfigure}
    \centering
    \includegraphics[width=25cm, height=15cm]{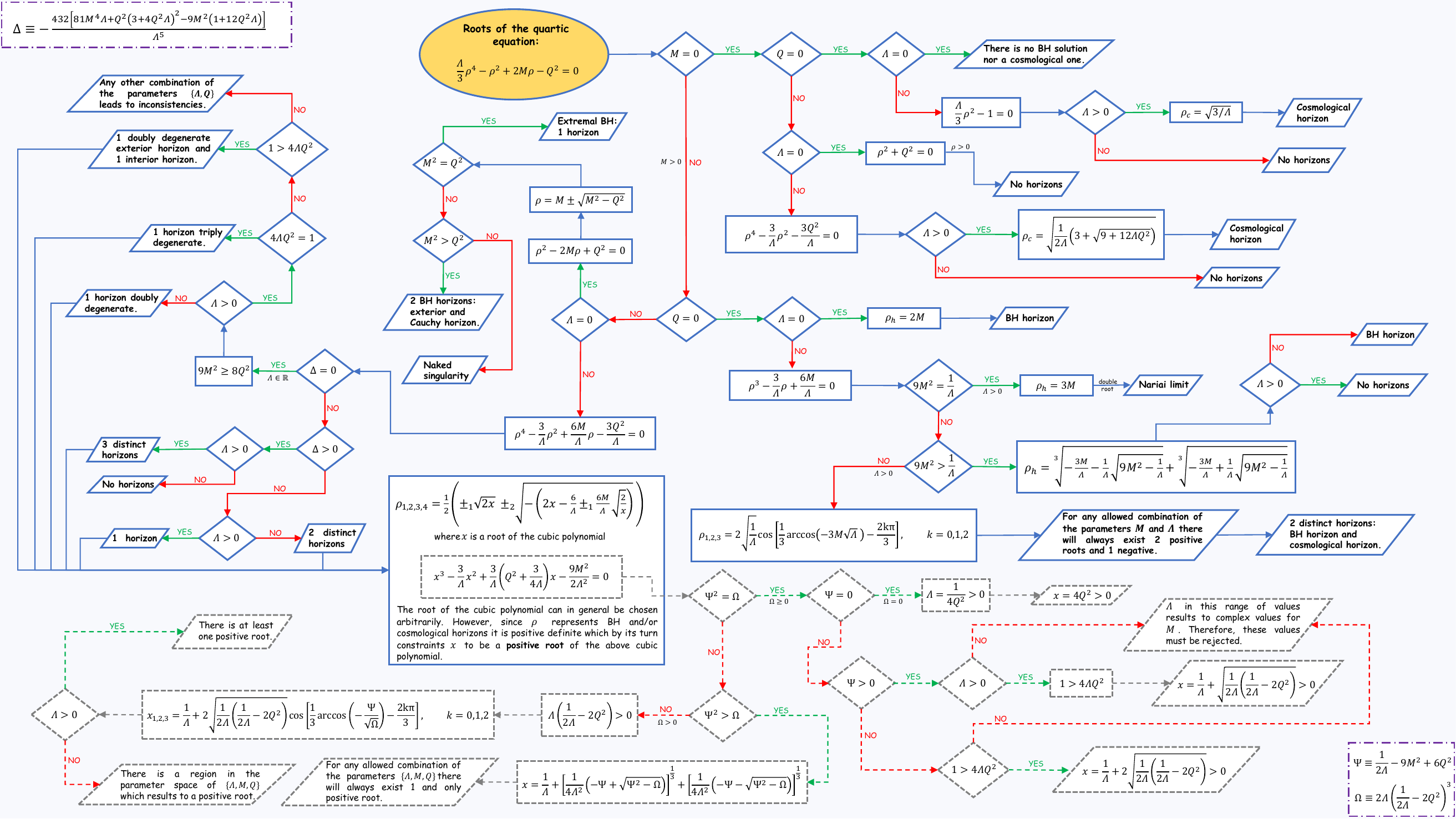}
    \caption{All possible roots of the quartic polynomial $f(\rho)=1-\frac{2M}{\rho}+\frac{Q^2}{\rho^2}-\frac{\Lambda}{3}\rho^2$.
   In the above flowchart, we catalogue the maximum possible number of horizons for each particular case.}
    \label{Hor-chart}
\end{sidewaysfigure}

In Fig. \ref{Hor-chart}, we depict a flowchart\footnote[3]{A flowchart is a graphical representation of a process
or a flow of consecutive steps. It was originated from computer science as a tool for representing algorithms and programming logic 
but nowadays plays an extremely useful role in displaying information visually and plainly. It is often the case that different 
flowcharts use different conventions about their symbols, thus, in our case we clarify that:
\begin{itemize}
\item \textbf{Ellipse/Terminator} represents the starting or ending point of the system.
\item \textbf{Rectangle/Process} represents a particular process, or a statement which is \textit{true}.
\item \textbf{Rhombus/Decision} represents a decision or a branching point. Lines coming out from the rhombus 
indicates different possible situations, leading to different sub-processes/sub-cases.
\item \textbf{Parallelogram/Data} represents information entering or leaving the system (input or output). In our
case it has mainly used as the final result/conclusion of each sub-case.
\end{itemize} } which constitutes an attentive 
scrutiny of the roots of the quartic polynomial  $f(\rho)=1-\frac{2M}{\rho}+\frac{Q^2}{\rho^2}-\frac{\Lambda}{3}\rho^2$. Every real, positive root
of this polynomial  corresponds either to a black-hole or a cosmological horizon of the five-dimensional 
spacetime \eqref{5d-metr}.  Let us consider some indicative cases. For $M \neq 0$ but $Q=0$ and $\Lambda=0$, 
we obtain the case of a five-dimensional spacetime with a sole black-hole horizon at $\rho_H=2M$. This
may be written in terms of $(r,y)$ as
\eq$\label{hor-loc-Scwarz}
r_H^2=4M^2-\frac{\big(e^{k|y_0|}-1\big)^2}{k^2}\,.$
This case was studied in \cite{NK1} where it was shown that the black-hole horizon was exponentially
localised close to the brane. Indeed, the aforementioned equation reveals the exponential decrease of $r_H$ as
$|y_0|$ increases and the existence of a value where the horizon vanishes, namely at $|y_0|=\ln (2M k+1)/k$.
Beyond this point, any $y$-slice of the five-dimensional spacetime is horizon-free and almost pure AdS, as
was shown in \cite{NK1}. In addition, the black-hole singularity was strictly localised on the brane as in the present
analysis.

Does the horizon exponential localisation persist also in the case of multiple horizons? Let us consider the case
with $M \neq 0$ and $Q \neq 0$ ($M^2>Q^2$), but $\Lambda=0$ for simplicity. In that case, it is easy to see that two horizons
emerge, an internal Cauchy horizon and an external event horizon located at $\rho_\pm= M \pm \sqrt{M^2-Q^2}$. 
Employing again the $(r,y)$ coordinates, these are re-written as
\eq$\label{hor-loc-Reissn}
r_\pm^2=\left(M \pm \sqrt{M^2-Q^2}\right)^2-\frac{\big(e^{k|y_0|}-1\big)^2}{k^2}\,.$
We observe that both horizons shrink as we move to $y$-slices of the bulk spacetime located further away from the brane.
Once again both horizons cease to exist beyond a certain value of $y$, namely at the values
\eq$\label{hor-loc-Reissn1}
|y_0|_{\pm}=\frac{1}{k} \,\ln\left[1+ kM \left(1 \pm \sqrt{1-\frac{Q^2}{M^2}}\right)\right]\,.$
Note that each horizon will vanish at its own value of the $y$-coordinate and that the horizon corresponding to
the smaller value of the radial coordinate $\rho$, i.e to the smaller root of the equation $f(\rho)=0$, will vanish first.

\begin{figure}[t!]
    \centering
  \hspace*{1cm}  \begin{subfigure}[b]{0.4\textwidth}
        \includegraphics[height=9cm]{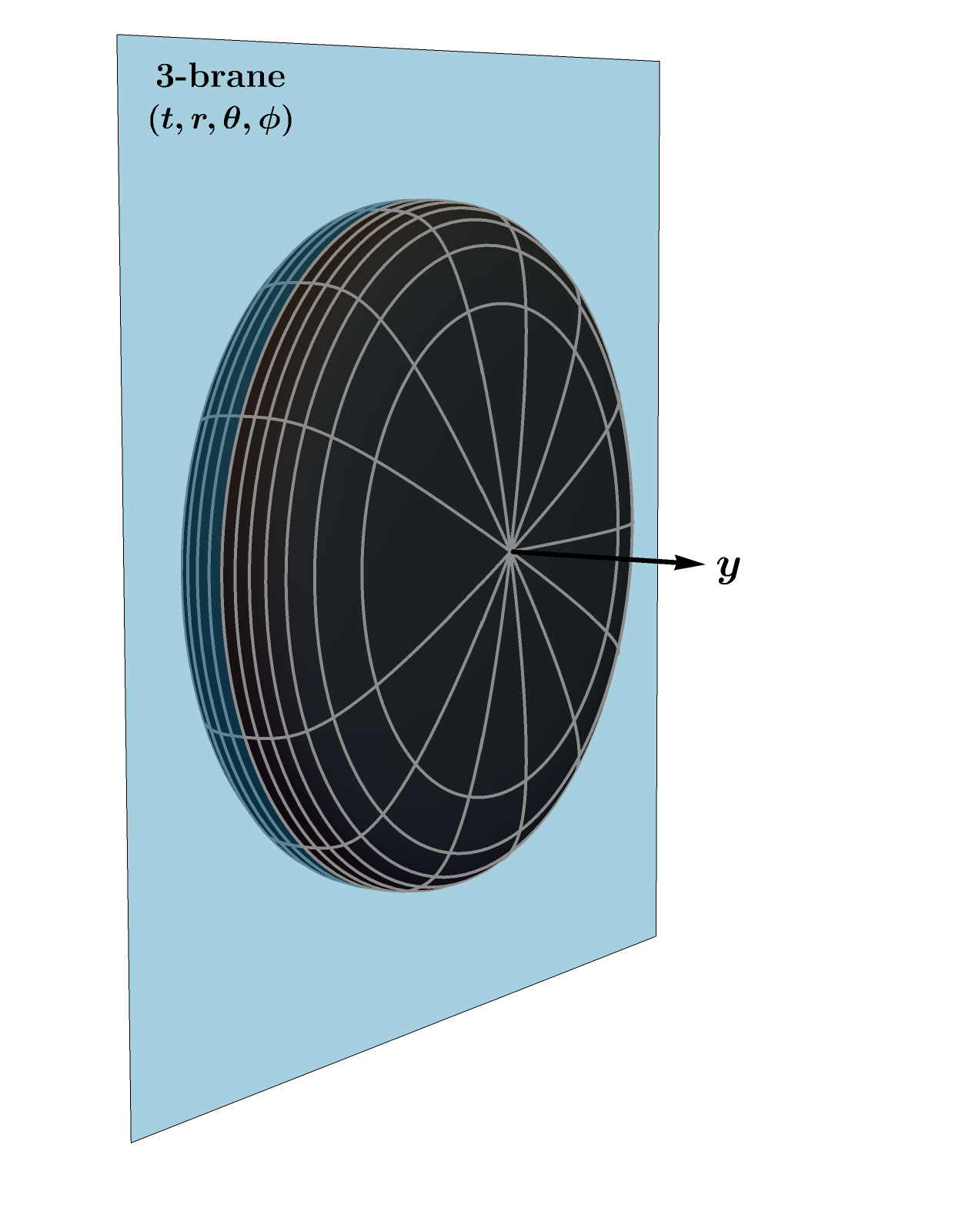}
        \caption{\hspace*{0em}}
        \label{loc-BH1a}
    \end{subfigure}
    ~ 
   \hspace*{0.5cm}  \begin{subfigure}[b]{0.4\textwidth}
        \includegraphics[height=9cm]{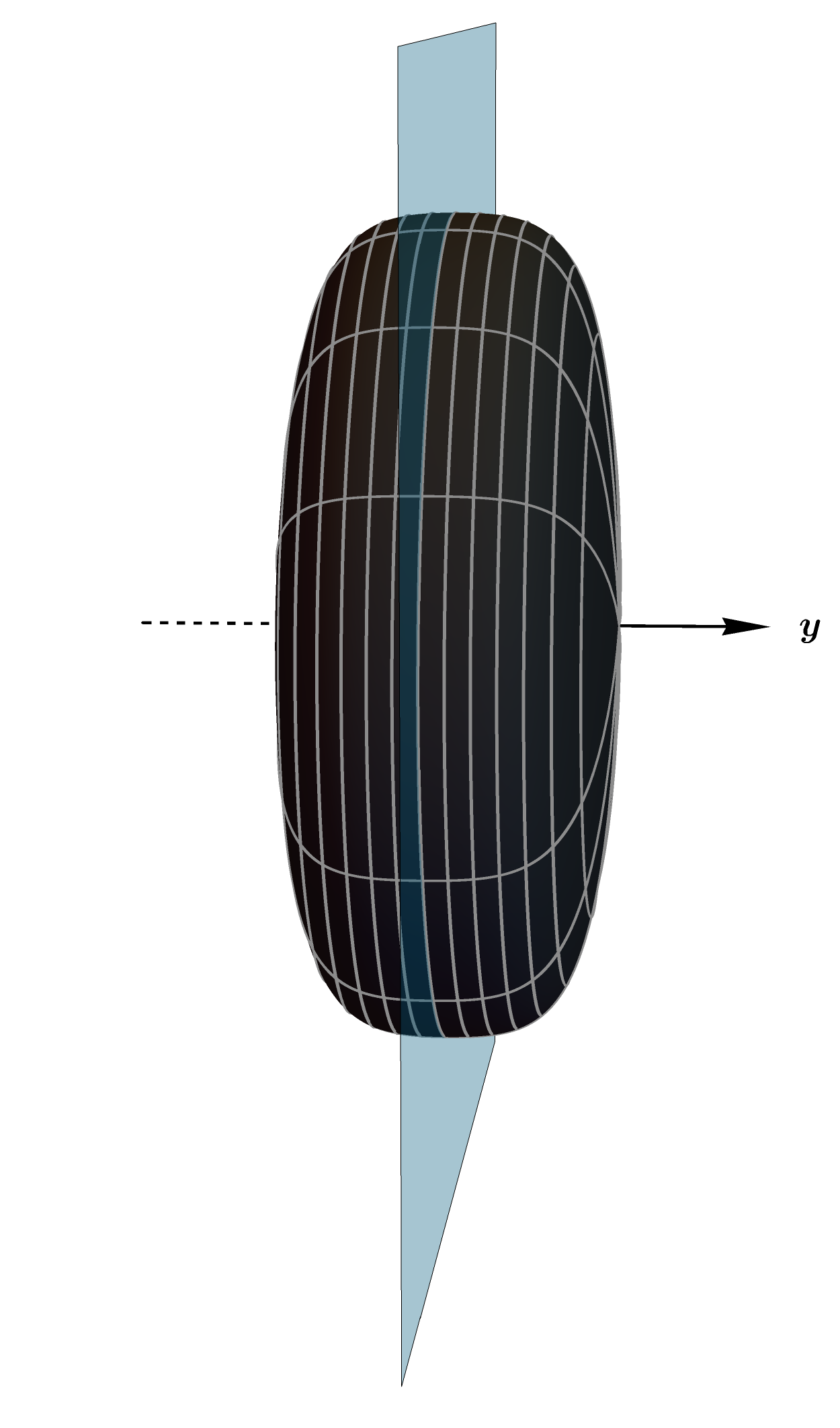}
        \caption{\hspace*{2.5em}}
        \label{loc-BH1b}
    \end{subfigure}  
    \vspace{-0.5em}  
    \caption{The horizon of the localized five-dimensional Schwarzschild black-hole from the bulk point of view for
    $M=7$, $k=0.5$, $Q=0$ and $\Lambda=0$. Both figures (a) and (b) depict the same image from different 
    angles. The depicted brane coordinates are the radial coordinate $r$ and the angular coordinate $\phi$.}
    \label{loc-BH-plot}
\end{figure}

The most general case arises when $M \neq 0$, $Q \neq 0$ and $\Lambda>0$. Then, we can have at most
three real, positive roots of the equation $f(\rho)=0$, and thus three horizons: an internal Cauchy horizon $\rho_-$,
an external event horizon $\rho_+$ and a cosmological horizon $\rho_{\ssst{C}}$. Their location in terms of the radial coordinate
$\rho$ is determined solely by the parameters $M$, $Q$ and $\Lambda$ and they naturally extend both on the brane
and in the bulk. As above, their profile in the bulk may be studied if we change to the $(r,y)$ coordinates; then,
the following general relation holds
\eq$\label{hor-loc}
r_h^2=\rho_h^2-\frac{\big(e^{k|y_0|}-1\big)^2}{k^2}\,,$
where the subscript $h$ has been used to denote the location of all three horizons. Since the value of $\rho_h$
is fixed by $M$, $Q$ and $\Lambda$, it is obvious that as $|y_0|$ increases the value of $r_h$ exponentially
decreases. Thus, as we move along the extra dimension away from the brane, $r_h$ quickly shrinks and becomes
zero at a distance
\eq$
|y_0|_{(h)}=\frac{1}{k}\ln(k\rho_h+1)\,.$
Again, each horizon will vanish at a different point along the extra dimension: the Cauchy horizon will vanish first,
the event horizon will follow next and the cosmological horizon will disappear last.\footnote[4]{The vanishing of the
cosmological horizon does not mean that the causal spacetime disappears but rather that a change of coordinates
is necessary (see Appendix \ref{app: CHCS}). After this point, a static ``observer'' no longer exists and a set of planar coordinates, such as
the ones used in cosmology to describe a time-depending de Sitter universe, is more appropriate. If one insists
in keeping the static, spherically-symmetric set of coordinates of Eq. \eqref{5d-metr}, and thus the notion of
a static ``observer'', then an interesting bound arises as to how far from the first brane a second one may be
introduced.}
Due to the exponential fall-off of each $r_h$ in terms of the $y$-coordinate, all horizons acquire a ``pancake''
shape with its longer dimension lying along the brane and its shorter one along the bulk. As an indicative case,
in Fig. \ref{loc-BH-plot} we give the geometrical representation of the event horizon of the five-dimensional
Schwarzschild spacetime. It is important to stress that by introducing non-vanishing $Q$ or $\Lambda$ 
the depicted general behaviour does not change.

\section{The Gravitational Theory}
\label{sec: GT}

After constructing the geometrical set-up of our five-dimensional gravitational theory, we now consider its action functional
which is described by the general expression
\eq$\label{bulk-action}
S_{B}=\int d^5x\, \sqrt{-g}\left(\frac{R}{2\kappa_5^2}+\lagr^{(B)}_{m}\right).$
In the above, $R$ is the Ricci scalar constructed in terms of the metric tensor $g_{MN}$ of the five-dimensional
spacetime, and $\kappa_5^2=8\pi G_5$  incorporates the five-dimensional gravitational constant $G_5$. All matter
fields, which may exist in the bulk, are described by the Lagrangian  density $\lagr^{(B)}_m$. Notice that we have
not explicitly included the five-dimensional cosmological  constant $\Lambda_5$ in the above action --  as we will
see in the analysis below, this quantity will appear naturally inside the bulk energy-momentum tensor. 

\par By varying the aforementioned action functional $S_B$ with respect to the metric tensor $g_{MN}$, we may
derive the gravitational field equations in the bulk which have the form
\eq$\label{field-eqs}
G_{MN}=\kappa_5^2\, T^{(B)}_{MN}\,.$
The quantity  $G_{MN}=R_{MN}-\frac{1}{2}\,g_{MN}R$ denotes the Einstein-tensor while
\eq$T^{(B)}_{MN}=-\frac{2}{\sqrt{-g}}\frac{\delta\left(\lagr^{(B)}_{m}\sqrt{-g}\right)}{\delta g^{MN}}$
is the bulk energy-momentum tensor associated with the Lagrangian density $\lagr^{(B)}_m$.  If we use the 
gravitational background \eqref{5d-metr} constructed in the previous section, we find that the 
non-zero components of $T^{(B)}_{MN}$ in mixed form are the following:
\gat$\label{rho-p1}
T^{(B)t}{}_t=T^{(B)\rho}{}_\rho=\frac{1}{\kappa_5^2}\left[2(3 k^2-\Lambda)+\frac{3 k\cos \chi }{\rho ^2}\left(3M-\frac{2Q^2}{\rho}\right)-\frac{3 M}{\rho ^3}\right],\\[2mm]
\label{p2}
T^{(B)\chi}{}_\chi=T^{(B)\theta}{}_\theta=T^{(B)\phi}{}_\phi=\frac{1}{\kappa_5^2}\left[2(3 k^2-\Lambda)-\frac{6k^2\cos^2\chi}\rho\left(M-\frac{Q^2}{\rho}\right)+\frac{6 k M \cos \chi }{\rho ^2}\right].$
The bulk energy-momentum tensor is thus characterized solely by three components: the energy-density
$\rho_E \equiv -T^{(B)t}{}_{t}$, the radial pressure $p_1 \equiv T^{(B)\rho}{}_{\rho}$,  and a common tangential
pressure $p_2 \equiv T^{(B)\chi}{}_{\chi}=T^{(B)\theta}{}_{\theta}= T^{(B)\phi}{}_{\phi}$. Therefore, the
gravitational background \eqref{5d-metr} of a five-dimensional, localised close to the brane black-hole solution
may be supported by a {\it diagonal} energy-momentum tensor which describes an anisotropic fluid. Employing
the fluid's timelike five-velocity $U^M$ and a spacelike unit vector in the direction of $\rho$-coordinate satisfying
the relations 
\gat$\label{U-vec}
U^M=\{U^t,0,0,0,0\},\hspace{1em} U^M U^N g_{MN}=-1\,,\\[2mm]
\label{X-vec}
X^M=\{0,X^\rho,0,0,0\},\hspace{1em} X^M X^N g_{MN}=1\,,$
the bulk energy-momentum tensor may be written in a covariant notation as follows
\eq$\label{en-mom}
T^{(B)MN}=(\rho_E+p_2)U^M U^N+(p_1-p_2)X^M X^N+p_2\, g^{MN}\,.$
The aforementioned, rather minimal, content of the bulk energy-momentum tensor was first found in the case
where the brane background was assumed to be the Schwarzschild solution \cite{NK1}. As we see, this structure
persists also in the case where the brane background assumes the form of more generalised four-dimensional black-hole
solutions. 

In fact, there are only two independent components of the energy-momentum tensor, namely the energy-density
$\rho_E$ and the tangential pressure component $p_2$; as Eq. (\ref{rho-p1}) reveals, the radial pressure component
$p_1$ is found to satisfy the equation of state $p_1=-\rho_E$ everywhere in the bulk. In addition, at asymptotic
infinity, i.e. as $\rho\ra+\infty$,  all three components of the energy-momentum tensor reduce to a constant
value, which can be identified as the five-dimensional cosmological constant $\Lambda_5$,
\gat$\label{as-rho}
\lim_{\rho\ra+\infty}\rho_{E}(\rho,\chi)=-\frac{2(3k^2-\Lambda)}{\kappa_5^2} \equiv \Lambda_5\,, \\[1mm]
\label{as-p2}
\lim_{\rho\ra+\infty}p_1(\rho,\chi)=\lim_{\rho\ra+\infty}p_2(\rho,\chi)=\frac{2(3k^2-\Lambda)}{\kappa_5^2} \equiv
-\Lambda_5\,.$
As discussed in the previous section and also confirm here, the asymptotic
form of the bulk spacetime depends on the sign of the quantity $(3k^2-\Lambda)$. If $3k^2>\Lambda$, the asymptotic
form of the energy-momentum tensor reduces to that of a negative bulk cosmological constant and the curvature invariant
quantities $R$, $\mathcal{R}$, $\mathcal{K}$ match the ones of an AdS$_5$ spacetime. Then, the brane parameter
$\Lambda$ is determined through the relation $2\Lambda=6k^2 -\kappa^2_5 |\Lambda_5|$, and its exact value depends
on the inter-balance between the warp parameter $k$ and the bulk cosmological constant $\Lambda_5$. In the special
case where a fine-tuning is imposed so that $\Lambda=0$,  the metric \eqref{5d-metr} incorporates exactly the 
Randall-Sundrum model \cite{RS1, RS2} at the spacetime boundary. Note, however, that the form of the warp factor
remains of an exponential form, i.e. $e^{-k |y|}$,  in our analysis regardless of the value of $\Lambda$. 

\begin{figure}[t!]
    \centering
    \begin{subfigure}[b]{0.485\textwidth}
    	\includegraphics[width=\textwidth]{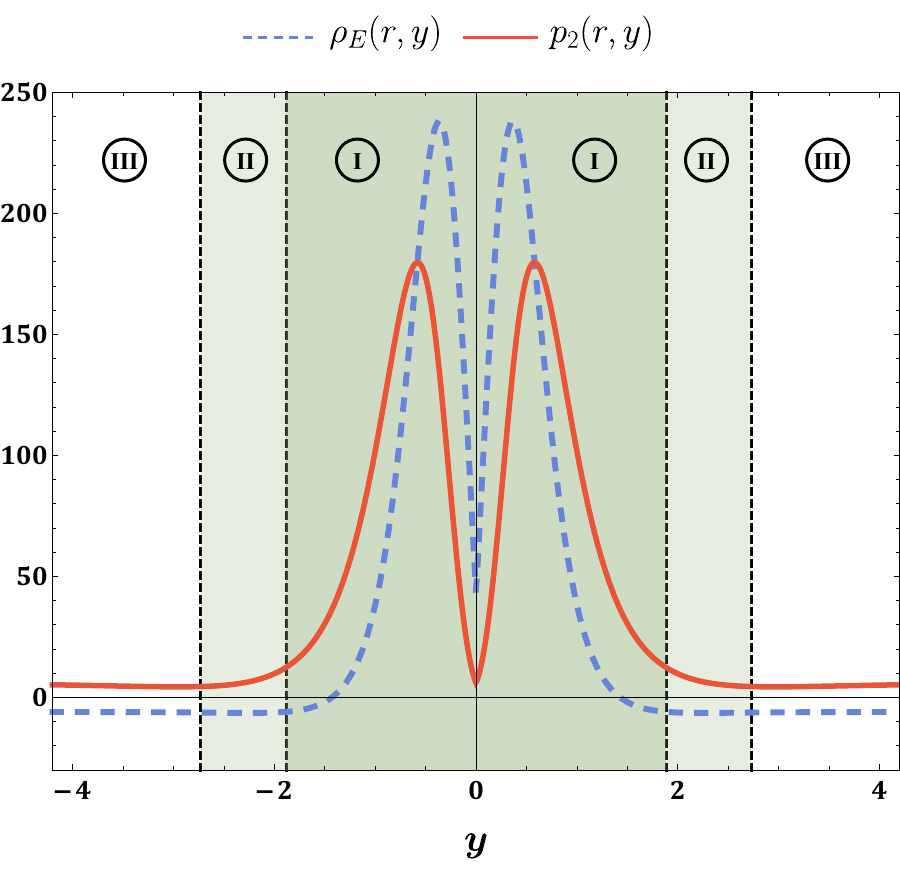}
    	\caption{\hspace*{-1.8em}}
    	\label{rho-p2-plot}
    \end{subfigure}
    \hfill
    \begin{subfigure}[b]{0.5\textwidth}
    	\includegraphics[width=\textwidth]{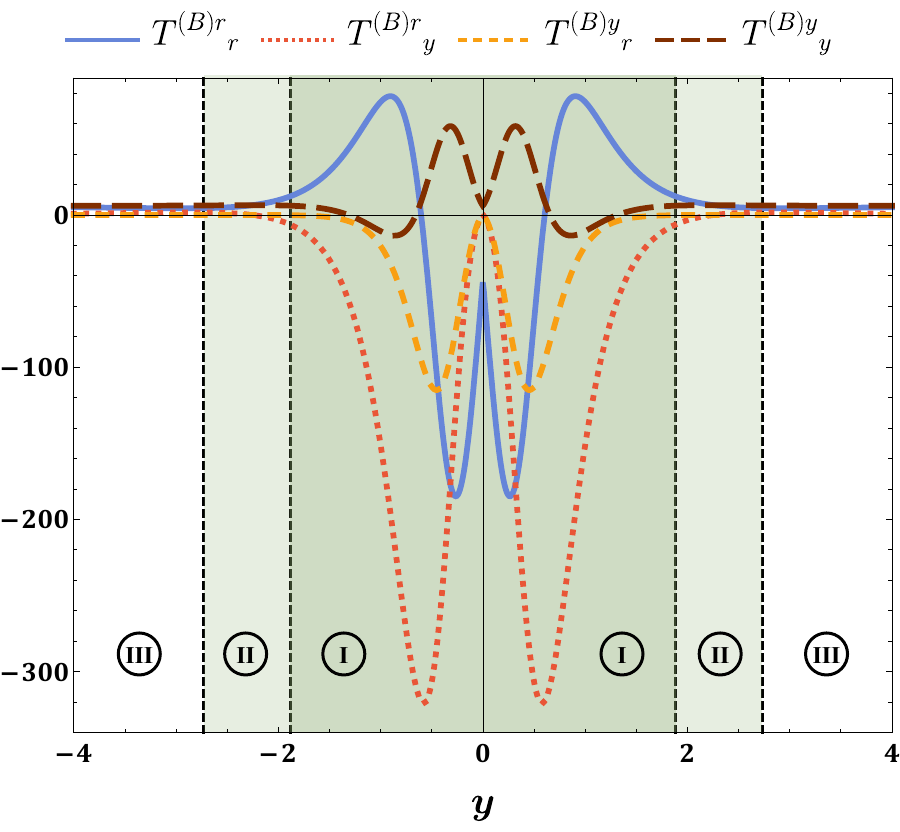}
		\caption{\hspace*{-2.3em}}    	
    	\label{Ene-Mom-comp-plot}
    \end{subfigure}
    \caption{(a) The profiles of the energy density $\rho_E$ and tangential pressure $p_2$ in terms of the $y$-coordinate
     for $\kappa_5=1$, $k=1$, $M=10$, $Q=9$, $\Lambda=5\times 10^{-21}$ and $r=0.85$. (b) The profiles of $T^{(B)r}{}_r$,
     $T^{(B)r}{}_y$, $T^{(B)y}{}_r$ and $T^{(B)y}{}_y$ depicted for the same values of the parameters.  Region I lies inside
     the Cauchy horizon, region II corresponds to the bulk spacetime between the two black-hole horizons, while
     region III is  located between the exterior black-hole horizon and the cosmological one.}
     \label{EneMom-plot}
\end{figure}

In order to study in more detail the profiles of the energy density $\rho_E$ and pressure $p_2$ in the bulk, we employ again the 
coordinates $(r,y)$. Using Eqs. \eqref{new-coords-inv}, \eqref{rho-p1} and \eqref{p2}, we find
\gat$\label{rho-ry}
\rho_E(r,y)=-\frac{1}{\kappa_5^2}\left\{2(3k^2-\Lambda)-\frac{3Mk^3\left(4-3e^{k|y|}\right)}{\left[k^2r^2+\left(e^{k|y|}-1\right)^2\right]^{3/2}}
-\frac{6Q^2k^4\left(e^{k|y|}-1\right)}{\left[k^2r^2+\left(e^{k|y|}-1\right)^2\right]^{2}} \right\}\,,\\[2mm]
\label{p2-ry}
p_2(r,y)=\frac{1}{\kappa_5^2}\left\{2(3k^2-\Lambda)+\frac{6Mk^3\left(e^{k|y|}-1\right)
\left(2-e^{k|y|}\right)}{\left[k^2r^2+\left(e^{k|y|}-1\right)^2\right]^{3/2}}+\frac{6Q^2k^4\left(e^{k|y|}-1\right)^2}{\left[k^2r^2
+\left(e^{k|y|}-1\right)^2\right]^{2}} \right\}\,.$
In Fig. \myref{EneMom-plot}{rho-p2-plot}, we present the profiles of the energy density $\rho_E$ and tangential pressure $p_2$ in terms of the
bulk coordinate $y$. In this indicative case, the values of the parameters were chosen to be $\kappa_5=1$, $k=1$, $M=10$,
$Q=9$, $\Lambda=5\times 10^{-21}$, and we have also fixed the radial coordinate on the brane at the random value $r=0.85$.
Substituting the aforementioned values of $M$, $Q$ and $\Lambda$ in the flowchart of Fig. \ref{Hor-chart}, one
can evaluate the locations of the three distinct horizons, namely $\rho_{-}=5.64$ (Cauchy horizon), $\rho_{+}=14.36$ (exterior 
black-hole horizon) and $\rho_{\ssst{C}}=2.45 \times 10^{10}$ (cosmological horizon). Given these values and the fixed radial
distance $r=0.85$, it is straightforward to calculate from Eq. \eqref{hor-loc} the corresponding values of $y$ at
which we encounter the three horizons in the bulk: the Cauchy horizon lies at $y_{-}=1.88$, the exterior black-hole horizon at 
$y_{+}=2.73$, and the cosmological horizon at $y_{c}=23.92$. We denote the bulk region inside the Cauchy horizon as region I,
the region between the two black-hole horizons as region II, and the region between the exterior black-hole horizon and the
cosmological horizon as region III; we denote these regions also in Fig. \myref{EneMom-plot}{Ene-Mom-comp-plot}. 

We observe that both the energy density $\rho_E$ and tangential pressure $p_2$ exhibit a shell-like distribution in region I,
i.e. in the region between the brane, located at $y=0$, and the Cauchy horizon. As the latter is approached, both components
quickly decrease towards their AdS$_5$ asymptotic values given by Eqs. \eqref{as-rho} and \eqref{as-p2}. These values are
adopted even before the exterior black-hole horizon is reached, therefore, as Fig. \myref{EneMom-plot}{rho-p2-plot} clearly depicts, region III
describes a pure AdS$_5$ spacetime. On the brane, the energy density $\rho_E$ and tangential pressure $p_2$ adopt
values which respect all energy conditions since there we have $\rho_E>0$, $\rho_E+p_1=0$ and $\rho_E>p_2$. Although the
profiles of $\rho_E$ and $p_2$ depend on the chosen values of the parameters of the theory, the behaviour depicted in 
Fig. \myref{EneMom-plot}{rho-p2-plot} is by no means a special one and in fact arises for a large number of sets of parameter values. What we
should also stress is the emergence of a regime close to the Cauchy horizon where the energy conditions are violated
since $\rho_E < p_2$. The same behaviour was also observed in our previous work \cite{NK1} and seems to be a requisite
for the localisation of the black-hole topology close to the brane as well as for the transition to a pure AdS$_5$ spacetime which, by
construction, is characterised by the relation $p_i=-\rho_E=|\Lambda_5|$.

 When we perform the coordinate change described via Eq. \eqref{new-coords-inv}, the components $T^{(B)M}{}_N$ of the
 energy-momentum tensor are bound to change. The $T^{(B)t}{}_t \equiv -\rho_E$ and $T^{(B)\theta}{}_\theta=T^{(B)\phi}{}_\phi \equiv p_2$
 components receive no additive corrections and their change amounts to merely substituting $\{\rho, \chi\}$ by $\{r,y\}$ in
 their expressions, thus leading to Eqs. \eqref{rho-ry} and \eqref{p2-ry}.
 However, the $T^{(B)r}{}_r$, $T^{(B)r}{}_y$, $T^{(B)y}{}_r$ and $T^{(B)y}{}_y$ components receive also additive corrections and their
 expressions are significantly modified. The analysis leading to the new expressions of all the components of the energy-momentum
 tensor is given in Appendix \ref{app: En-Mom}.  Therefore, for completeness, in Fig. \myref{EneMom-plot}{Ene-Mom-comp-plot} we depict also the behaviour of
 these four components of the energy-momentum tensor in terms of  the extra dimension $y$ and for the same set of parameter
values as in Fig.  \myref{EneMom-plot}{rho-p2-plot}.  As was the case with $\rho_E$ and $p_2$, these components remain everywhere regular,
have a shell-like distribution inside region I and quickly adopt their asymptotic values even before the exterior black-hole
horizon is reached: for the $T^{(B)r}{}_r$ and $T^{(B)y}{}_y$ components, this asymptotic value is $|\Lambda_5|$ while,
for the off-diagonal components  $T^{(B)r}{}_y$ and $T^{(B)y}{}_r$, this asymptotic value is zero, as expected.
Note also that the coordinate change \eqref{new-coords-inv} destroys the simple relations \eqref{rho-p1}, \eqref{p2}
between the components of the energy momentum tensor which were
valid in the $\{t,\rho,\chi,\theta,\phi\}$ set of coordinates. All these reveal that, although the ``axial''  set of coordinates
$\{t,r,\theta,\phi,y\}$ serve better to illustrate the behaviour of both the curvature and distribution of matter {\it with respect
to the brane observer}, it is the ``spherical'' set of coordinates $\{t,\rho,\chi,\theta,\phi\}$ which {\it encodes the highest
symmetry} of the five-dimensional theory and leads to the simplest profile of both the spacetime and the energy-momentum tensor. 


\smallskip \medskip
\subsection{A Field-Theory Toy-Model}

\par In this subsection, we will investigate further the nature of the bulk energy-momentum tensor  which is
necessary  to support the geometry of the five-dimensional localized  Reissner-Nordstr\"om-(A)dS black hole
presented in Section \ref{sec: GS}.  Due to the simple structure of $T_{MN}^{(B)}$, given in Eqs. (\ref{rho-p1}), (\ref{p2}),
the term ``anisotropic fluid'' was used to describe it, and a covariant form for its expression was also found.
However, it would be interesting to see if a field-theory model could be proposed to support it, and under
which conditions on the associated fields this task could be fulfilled. 

In the following analysis, we will use the spherically-symmetric set of coordinates  $\{t,\rho,\chi,\theta,\phi\}$
in which $T^{(B)}_{MN}$ takes its simplest possible form, as argued above. We will employ five-dimensional
fields that are allowed to propagate outside our brane, and thus consider scalar or gauge fields which are
distinct from the corresponding Standard-Model degrees of freedom living on the brane. According to our
results, a theory with only minimally-coupled scalars or with only minimally-coupled vector fields fails to lead
to the desired structure of the bulk energy-momentum tensor.  We therefore consider a tensor-vector-scalar
five-dimensional field theory where the bulk matter Lagrangian density $\lagr^{(B)}_{m}$ appearing in
Eq. (\ref{bulk-action}) is given by  
\gat$\label{lagr} 
\lagr^{(B)}_{m}:= \lagr^{(g)}+\lagr^{(sc)}\,,$
with
\gat$
\label{lagr-em}
\lagr^{(g)}:=-\frac{1}{4}\,F^{MN}F_{MN}\,,\\[2mm]
\label{lagr-sc}
\lagr^{(sc)}:=-f_1(\xi,\psi)(\pa\xi)^2-f_2(\xi,\psi)(\pa \psi)^2-V(\xi,\psi)\,.$
Above, $F_{MN}=\nabla_M A_N - \nabla_N A_M$ is the field-strength tensor of an Abelian gauge field $A_M$ and 
$\{\xi(\rho,\chi),\psi(\rho,\chi)\}$ are two scalar fields. In addition, we have introduced two arbitrary functions
$f_1(\xi,\psi)$ and $f_2(\xi,\psi)$ in the kinetic terms of the scalar fields as well as an interaction potential $V(\xi,\psi)$.
The variation of $\sqrt{-g}\,\lagr^{(B)}_{m}$ with respect to $g_{MN}$ leads to the result
\eq$
T^{(B)MN} = T^{(g)MN} + T^{(sc)MN}\,,$
where
\eq$\label{ene-mom-em}
T^{(g)MN} =F^{MA}F^N{}_A-\frac{1}{4}\,g^{MN}F^{AB}F_{AB}\,,$
and
\eq$\label{ene-mom-sc}
T^{(sc)}_{MN}=2f_1(\xi,\psi)\pa_M\xi\,\pa_N\xi+2f_2(\xi,\psi)\pa_M\psi\,\pa_N\psi+g_{MN}\,\lagr^{(sc)}\,.$

In what follows, we will also assume the following configuration for the gauge field-strength tensor
\eq$\label{em-tensor}
\left(F^{MN}\right)=\left(\begin{array}{ccccc}
0 & E_1(\rho,\chi) & E_2(\rho,\chi) & 0 & 0\\[1mm]
-E_1(\rho,\chi) & 0 & 0 & 0 & 0\\[1mm]
-E_2(\rho,\chi) & 0 & 0 & 0 & 0\\[1mm]
0 & 0 & 0 & 0 & B(\rho,\chi,\theta)\\[1mm]
0 & 0 & 0 & -B(\rho,\chi,\theta) & 0
\end{array}\right),$
where $E_1$, $E_2$ and $B$ stand for two components of the ``electric'' bulk gauge field and a sole
component of the ``magnetic'' field, respectively. Employing the above in Eq. (\ref{ene-mom-em}), one may
easily calculate the components of the gauge-field energy-momentum tensor $T^{(g)MN} $. Using also
the expression (\ref{ene-mom-sc}), the components of $T^{(sc)}_{MN}$ for the two scalar fields readily
follow. Taking their sum, we obtain the following results for the non-vanishing mixed components of the bulk
energy-momentum tensor 
\begin{eqnarray}
& T^{(B)t}{}_t=\frac{1}{2}(-b B^2+a_1 E_1^2+a_2 E_2^2)-f_1(\pa\xi)^2-f_2(\pa\psi)^2-V,&  \nonumber \\[3mm]
 & T^{(B)\rho}{}_\rho=\frac{1}{2}(-b B^2+a_1 E_1^2-a_2 E_2^2)+f_1\left(\pa^\rho\xi\pa_\rho\xi-\pa^\chi\xi\pa_\chi\xi\right)
+f_2\left(\pa^\rho\psi\pa_\rho\psi-\pa^\chi\psi\pa_\chi\psi\right)-V,& \nonumber\\[3mm]
& T^{(B)\rho}{}_\chi = a_2E_1 E_2+ 2\left(f_1\,\pa^\rho\xi\pa_\chi\xi+f_2\,\pa^\rho\psi\pa_\chi\psi\right),& \label{total-emt}\\[3mm]
& T^{(B)\chi}{}_\chi=\frac{1}{2}(-b B^2-a_1 E_1^2+a_2 E_2^2)-f_1\left(\pa^\rho\xi\pa_\rho\xi-\pa^\chi\xi\pa_\chi\xi
\right)-f_2\left(\pa^\rho\psi\pa_\rho\psi-\pa^\chi\psi\pa_\chi\psi\right)-V, &\nonumber\\[3mm]
& T^{(B)\theta}{}_\theta=T^{(B)\phi}{}_\phi=\frac{1}{2}(b B^2-a_1 E_1^2-a_2 E_2^2)-f_1(\pa\xi)^2-f_2(\pa\psi)^2-V. &
\nonumber
\end{eqnarray}
In the above, we have defined the quantities $a_1\equiv g_{tt}\,g_{\rho\rho}$, $a_2\equiv g_{tt}\,g_{\chi\chi}$ and 
$b\equiv g_{\theta\theta}\,g_{\phi\phi}$ for simplicity.

Let us now investigate whether the above set of components can be simplified in order to resemble the minimal 
configuration described by Eqs. \eqref{rho-p1}, \eqref{p2}. We thus first demand that $T^{(B)t}{}_t= T^{(B)\rho}{}_\rho$,
and we obtain the constraint
\eq$\label{E2-eq}
E_2^2=\frac{2\left(f_1\,\pa^\rho\xi\pa_\rho\xi+f_2\,\pa^\rho\psi\pa_\rho\psi\right)}{a_2}\,.$
We next observe that the configuration of Eqs. \eqref{rho-p1}, \eqref{p2} has no off-diagonal component. Thus
demanding that $T^{(B)\rho}{}_\chi =0$ and employing Eq. \eqref{E2-eq}, we also obtain 
\eq$\label{E1-eq}
E_1^2=\frac{1}{a_2}\frac{2\left(f_1\,\pa^\rho\xi\pa_\chi\xi+f_2\,\pa^\rho\psi\pa_\chi\psi\right)^2}{f_1\,\pa^\rho\xi\pa_\rho\xi
+f_2\,\pa^\rho\psi\pa_\rho\psi}\,.$
Demanding finally that $T^{(B)\chi}{}_\chi= T^{(B)\theta}{}_\theta = T^{(B)\phi}{}_\phi$, we are led to a third constraint
\eq$\label{B-eq}
B^2=\frac{2\left[f_1(\pa\xi)^2+f_2(\pa\psi)^2\right]}{b}\,.$
Therefore, the co-existence in the bulk of the three components of the field-strength tensor with the two scalar fields,
in a way that they satisfy the above three constraints, ensures that the total energy-momentum tensor in the bulk acquires
the form dictated by Eqs. \eqref{rho-p1}, \eqref{p2}.  

In addition, setting $\rho_E \equiv -T^{(B)t}{}_t$ and $p_2 \equiv T^{(B)\chi}{}_\chi$, the remaining two components give 
\eq$\label{sc-lagr-eq}
f_1(\pa\xi)^2+f_2(\pa\psi)^2+V=\frac{1}{2}(\rho_E-p_2)\,,$
\eq$\label{em-eq}
bB^2-a_1 E_1^2-a_2 E_2^2=\rho_E+p_2\,.$
Therefore, the two independent components of the energy-momentum tensor in the bulk are determined by the
exact profiles of the gauge and scalar fields. These in turn must satisfy  their own equations of motion. By 
considering the variation of the action $S_B$ with respect to $A_M$, we obtain the five-dimensional equation
for the gauge field in the bulk, namely
\eq$\label{maxwell-eqs}
\pa_M\left(\sqrt{-g}\,F^{MN}\right)=0\,.$
Considering the components $N=t$ and $N=\phi$, we find
\bal$
\label{max-eq-1}
& \displaystyle{\pa_\rho E_1+\pa_\chi E_2+E_1\frac{\pa_\rho\sqrt{-g}}{\sqrt{-g}}+E_2\frac{\pa_\chi\sqrt{-g}}{\sqrt{-g}}=0}\,,\\[2mm]
\label{max-eq-2}
 & \pa_\theta B+\frac{B\,\cos\theta}{\sin\theta}=0\Ra B(\rho,\chi,\theta)=\frac{B_0(\rho,\chi)}{\sin\theta} \,,$
respectively, while the remaining components are identically zero. Additionally, the variation of $S_B$ with respect to the scalar
fields $\xi$ and $\psi$ results in the equations
\eq$\label{xi-eq}
\frac{1}{2}\left[(\pa_\xi f_1)(\pa\xi)^2+(\pa_\xi f_2)(\pa\psi)^2+\pa_\xi V\right]=\frac{\pa_M\left(\sqrt{-g}\,f_1\,g^{MN}\pa_N\xi
\right)}{\sqrt{-g}}\,,$
\eq$\label{psi-eq}
\frac{1}{2}\left[(\pa_\psi f_1)(\pa\xi)^2+(\pa_\psi f_2)(\pa\psi)^2+\pa_\psi V\right]=\frac{\pa_M\left(\sqrt{-g}\,f_2\,g^{MN}\pa_N\psi
\right)}{\sqrt{-g}}\,.$

The above set of four differential equations \eqref{max-eq-1}-\eqref{psi-eq}, together with the constraints \eqref{E1-eq}-\eqref{em-eq},
may indeed possess a mathematically consistent solution. The complexity of the system would most likely demand numerical
calculation for this solution to be derived. However, instead of attempting to solve this coupled system of equations, we would like to 
examine the ensuing characteristics of the fields. To this end, let us focus on Eq. \eqref{em-eq}: employing the exact form of the 
components $\rho_E$ and $p_2$ from Eqs. \eqref{rho-p1}, \eqref{p2}, we may rewrite it as
\eq$ \label{rho+p2}
bB^2-a_1 E_1^2-a_2 E_2^2=
\frac{1}{\kappa_5^2}\left[-\frac{6k^2\cos^2\chi}\rho\left(M-\frac{Q^2}{\rho}\right)-\frac{3 k\cos \chi }{\rho ^2}
\left(3M-\frac{2Q^2}{\rho}\right)+\frac{3 M}{\rho ^3}\right].$
The right-hand-side of the above equation is clearly not sign-definite. For small $\rho$, it is positive definite since
in this regime both $\rho_E$ and $p_2$ are positive, as Fig. \myref{EneMom-plot}{rho-p2-plot} reveals. However, as $\rho$
increases, negative-valued terms inside the square brackets begin to dominate making this combination  clearly
negative for large values of $\rho$. Since $a_1<0$, $a_2<0$ and $b>0$,  according to their definitions below
Eq. (\ref{total-emt}), this means that  at least one of the components of the gauge field strength-tensor
$F_{MN}$ must turn imaginary near the bulk boundary. Due to the constraints \eqref{E1-eq}, \eqref{B-eq},
this may lead to $\xi$ or $\psi$ also becoming imaginary. 

Simpler variants of the above model may also be built, however, they all suffer from the above problem. For instance,
if we consider the case with $E_2=0$ and $\xi=\xi(\chi)$ together with the condition $f_2=0$, the energy-momentum
tensor comes out to be automatically diagonal and satisfying $T^{(B)t}{}_t=T^{(B)\rho}{}_\rho$. The constraints
\eqref{E2-eq}, \eqref{E1-eq} now disappear while the one for $B$ still holds. The gauge-field equations 
\eqref{max-eq-1} and \eqref{max-eq-2} are easily satisfied for a wide range of choices for $E_1$ and $B$.
The second scalar field $\psi(\rho,\chi)$ is now an auxilary field whose equation of motion \eqref{psi-eq}
 introduces a constraint between $f_1$ and $V$. Nevertheless, Eq. \eqref{rho+p2} still holds with $E_2=0$,
 and thus the necessity for a ``phantom'' gauge field (and a ``phantom'' scalar field) at the bulk boundary still exists. 
 
 Phantom scalar fields are often used in the context of four-dimensional analyses as a mean to create the
 necessary yet peculiar dark energy component with $w < -1$ in our universe. In our analysis, a bulk matter
 with also peculiar characteristics seems to be necessary to localise a five-dimensional black hole on the brane,
 otherwise its singularity would leak in  the bulk.  The desired structure of the bulk energy-momentum tensor
 as well as the introduction of the ``charge'' parameter $Q$ in our metric demand the presence of gauge and
 scalar fields with phantom-like properties at the bulk boundary.  We should stress that all fields are ``ordinary'' 
 close  to our brane and no violation of energy conditions takes place on our brane. Could a gauge field, that
 turns phantom-like at the outskirts of the bulk spacetime, be considered as ``natural'' or at least acceptable?
 Such an analysis, although well-motivated, would take us beyond the scope of  the present study and is thus
 left for a future work.


\section{Junction Conditions and Effective Theory}
\label{sec: JC-ET}

In this final section, we turn our attention from the structure and content of the five-dimensional spacetime to issues
related to the presence of the brane itself, namely its consistent embedding in the bulk and the effective four-dimensional
gravitational equations. A detailed derivation of the effective theory on the 3-brane in brane-world models was presented
in \cite{SMS}, however, in order to keep our analysis self-contained, we will reproduce here the main results and equations.
It is also important to note that in \cite{SMS} the bulk matter of the brane-world model was described only by a negative
cosmological constant, whereas in our case the bulk spacetime contains an anisotropic fluid, a feature that slightly
modifies some parts of the analysis.

In the standard brane-world scenario, our 3-brane ($\Sigma$, $h_{MN}$) is embedded in the five-dimensional spacetime
($\mathcal{M}$, $g_{MN}$) at $y=0$. The induced metric on the brane is defined via the relation
$h_{MN}\equiv\left(g_{MN}\right)_{y=0}-n_{M}n_{N}$, where $n^M$ is the unit normal vector to the 3-brane. From
Eq. \eqref{metr-r-y}, we may deduce that $n^M=\del^M{}_y$. In what follows, we will denote tensors on $\Sigma$ with
a bar to be distinguished from the corresponding five-dimensional tensors.
Using the \textit{Gauss's Theorema Egregium}\footnote[5]{We note that the square brackets $[]$ in a tensor's indices denote 
anti-symmetrization, namely $A_{[MN]}\equiv\frac{1}{2}\left(A_{MN}-A_{NM}\right)$.}
\eq$\label{Gauss-eq}
\bar{R}^A{}_{BCD}=h^{A}{}_M\, h^{N}{}_B\,h^{K}{}_C\, h^{L}{}_D R^M{}_{NKL}+2K^{A}{}_{[C}K_{D]B}\,,$
and the \textit{Codazzi's equation}\footnote[6]{A tensor at a point $P\in\Sigma$ is invariant under the projection $h^{M}{}_N$ if
\eq$\label{tensor-on-brane}
T^{M_1M_2\cdots M_p}{}_{N_1N_2\cdots N_q}=h^{M_1}{}_{A_1}\,h^{M_2}{}_{A_1}\cdots h^{M_p}{}_{A_p}\,h^{B_1}{}_{N_1}
h^{B_2}{}_{N_2}\cdots h^{B_q}{}_{N_q}\,T^{A_1A_2\cdots A_p}{}_{B_1B_2\cdots B_q}\,.$
The covariant derivative $D_L$ on $\Sigma$ can be defined via the projection of the covariant derivative on $\mathcal{M}$; for any tensor
obeying \eqref{tensor-on-brane} we define 
\eq$\label{cov-der-on-brane}
D_{L}T^{M_1\cdots M_p}{}_{N_1\cdots N_q}=h^{K}{}_L\,h^{M_1}{}_{A_1}\cdots h^{M_p}{}_{A_p}\,h^{B_1}{}_{N_1}\cdots h^{B_q}{}_{N_q}\,
\nabla_KT^{A_1\cdots A_p}{}_{B_1\cdots B_q}\,.$}
\eq$\label{Codazzi-eq}
R_{AB}\,h^A{}_M\,n^B=D_L K^{L}{}_M-D_M\, K\,,$
we obtain the following relation for the Einstein tensor on the 3-brane:
\eq$\label{einstein-on-brane}
\bar{G}_{MN}=h^A{}_Mh^{B}{}_N\,G_{AB}+R_{AB}\,n^An^Bh_{MN}+KK_{MN}-K_{M}{}^LK_{LN}-\frac{1}{2}h_{MN}\left(K^2-
K^{AB}K_{AB}\right)-\wtild{E}_{MN}\,.$
In the above, $K_{MN}$ is the extrinsic curvature of the brane defined as
\eq$\label{extr-curv}
K_{MN}\equiv h^{A}{}_M\,h^{B}{}_N\,\nabla_A\,n_{B}=h^{L}{}_{M}\,\nabla_L\,n_{N}\,,$
while
\eq$\label{E-riem}
\wtild{E}_{MN}\equiv R^A{}_{BCD}\, n_A\,n^C\,h^B{}_M\,h^{D}{}_N\,.$
Decomposing the Riemann tensor into the Weyl curvature, the Ricci tensor and the Ricci scalar, we obtain
\eq$\label{riem-weyl}
R_{ABCD}=\frac{2}{3}\left(g_{A[C}R_{D]B}-g_{B[C}R_{D]A}\right)-\frac{1}{6}g_{A[C}\,g_{D]B}R+C_{ABCD}\,.$
Using the five-dimensional gravitational field equations \eqref{field-eqs} together with \eqref{riem-weyl} in \eqref{einstein-on-brane} we are led to
\gat$
\bar{G}_{MN}=\frac{2\kappa_5^2}{3}\left[h^A{}_Mh^{B}{}_N\,T^{(B)}_{AB}+\left(n^An^{B}\,T^{(B)}_{AB}-\frac{T^{(B)}}{4}\right)h_{MN}
\right]+KK_{MN}-K_{M}{}^LK_{LN}\nonum\\
\label{effective-theory}
-\frac{1}{2}h_{MN}\left(K^2-K^{AB}K_{AB}\right)-E_{MN}\,,$
where $T^{(B)}\equiv T^{(B)L}{}_L$ is the trace of the bulk energy-momentum tensor, and
\eq$\label{E-weyl}
E_{MN}\equiv C^A{}_{BCD}\, n_A\,n^C\,h^B{}_M\,h^{D}{}_N\,.$

As is usual in all brane-world scenarios, we may write the total energy-momentum tensor as the sum of the
bulk  $T^{(B)}_{MN}$ and brane $T^{(br)}_{\mu\nu}$ energy-momentum tensors, namely
\eq$\label{ene-mom-dec}
T_{MN}=T^{(B)}_{MN}+\del^{\mu}_M\del^{\nu}_N\,T^{(br)}_{\mu\nu}\del(y)\,.$
The brane  energy-momentum tensor can be decomposed further as follows
\eq$\label{brane-ene-dec}
T^{(br)}_{\mu\nu}=-\sigma\, h_{\mu\nu}+\tau_{\mu\nu}\,,$
where $\sigma$ is the tension of the brane, and $\tau_{\mu\nu}$ encodes all the
other possible sources of energy and/or pressure on the brane. A natural question which arises in the
context of our analysis is whether the consistent embedding of our brane in the five-dimensional
line-element (\ref{metr-r-y}) demands the introduction of a non-trivial $\tau_{\mu\nu}$ on the brane. 

In order to investigate this, we will study Israel's junction conditions \cite{Israel} at $y=0$. These require
that
\gat$
\label{jc1}
[h_{MN}]=0\,,\\[2mm]
\label{jc2}
[K_{\mu\nu}]=-\kappa_5^2\left(T^{(br)}_{\mu\nu}-\frac{1}{3}h_{\mu\nu}\,T^{(br)}\right)\,.$
In the above, the bracket notation for a quantity $X$ simply means 
\eq$\label{bracket-def}
[X]=\lim_{y\ra 0^+}X-\lim_{y\ra 0^-}X=X^{(+)}-X^{(-)}\,.$
Let us determine first the components of the induced metric on the brane $h_{MN}$. These are found to be
\eq$\label{ind-metr-comp}
(h_{MN})=\left(\begin{array}{ccccc}
-\left(1-\frac{2M}{r}+\frac{Q^2}{r^2}-\frac{\Lambda}{3}r^2\right) 		& 		0 		& 		0 		& 		0 		&   0		 \\[1mm]
0		 &		 \left(1-\frac{2M}{r}+\frac{Q^2}{r^2}-\frac{\Lambda}{3}r^2\right)^{-1} 		&		 0		 &		 0		& 0	  \\[1mm]
0		&		0		&		r^2		&		0	&	0	\\
0		&		0		&		0		&		r^2\sin^2\theta	&	0\\
0	&  0	&	0	&	0	&  0		
\end{array}\right)\,.$
We may easily see that they indeed satisfy Israel's first condition. Also, employing these, we may easily determine the
components of the extrinsic  curvature close to the 3-brane which have the form
\eq$\label{extr-curv2}
K_{MN}=-k \frac{d|y|}{dy}\,\del^{\mu}_M\del^{\nu}_N\, h_{\mu\nu}.$
The trace of $K_{MN}$ is also found to be $K=-4k\,(d|y|/dy)$. We may alternatively write Eq. \eqref{jc2}  as\footnote[7]{For an explicit
proof of Eq. \eqref{jc2-new} see Appendix \ref{app: Br-Ene}.}
\eq$\label{jc2-new}
T^{(br)}_{\mu\nu}=-\frac{1}{\kappa_5^2}\left([K_{\mu\nu}]-h_{\mu\nu}[K]\right)\,.$
Using Eq. (\ref{bracket-def}), the assumed $\mathbf{Z}_2$-symmetry of the model in the bulk
and the components of $K_{\mu\nu}$, we find 
\eq$\label{brane-ene-comp}
T^{(br)}_{\mu\nu}=-\frac{6k}{\kappa_5^2}\,h_{\mu\nu}\,.$
Comparing Eq. \eqref{brane-ene-comp} with Eq. \eqref{brane-ene-dec}, we easily deduce that
$\sigma=6k/\kappa_5^2>0$, while $\tau_{\mu\nu}=0$. This means that the consistent embedding of our
3-brane in the five-dimensional spacetime constructed in Section \ref{sec: GS}---and described by either the line-element
(\ref{metr-r-y}) or (\ref{5d-metr})---does not demand the introduction of any additional matter on the
brane\footnote[8]{The absence of the need for the introduction of any brane matter but the necessity for
the presence of bulk fields in order to localise the black-hole geometry close to the brane
could be related to similar conclusions derived following the effective-field-theory point-of-view in
braneworlds\cite{Fichet}.}. In the context of the five-dimensional theory, the brane
contains only  its constant positive self-energy $\sigma$.
In fact, it is this quantity together with the five-dimensional gravitational constant $\kappa_5^2$ that
determine the warp parameter $k$ of the line-element in the bulk. 

We may now proceed to derive the effective theory on the brane. The gravitational equations on the 3-brane
can be determined from Eq. \eqref{effective-theory} by setting $y=0$. We note that for either $M$ or $N$
equal to $y$, the r.h.s. of \eqref{effective-theory} is trivially zero; this implies that $\bar{G}_{yN}=0\ 
\forall N$, as expected. Due to the $\mathbf{Z}_2$-symmetry, we may perform the calculation either on
the $+$ or $-$ side of the brane, therefore we will omit the $\pm$ signs in what follows. Using the 
results for the induced metric $h_{MN}$, the extrinsic curvature $K_{MN}$ and
the normal vector $n^M$ derived above in \eqref{brane-ene-dec}, we obtain
\eq$\label{grav-eqs-br-new}
\bar{G}_{\mu\nu}=8\pi G_4\left(T^{(eff)}_{\mu\nu}+ \tau_{\mu\nu}\right)+\kappa_5^4\left(\pi_{\mu\nu}-\frac{\ \sig^2}{12}\,h_{\mu\nu}\right)
-E_{\mu\nu}\Big|_{y\ra 0}\,,$
where
\bal$
&G_4=\frac{\kappa_5^4\,\sigma}{48\pi}\,,\\[3mm]
&T^{(eff)}_{\mu\nu}\equiv\frac{2}{3k}\left[T^{(B)}_{\mu\nu}+\left(T^{(B)}_{yy}-\frac{T^{(B)}}{4}\right)h_{\mu\nu}\right]_{y=0}\,,\\[3mm]
&\pi_{\mu\nu}=-\frac{1}{4}\tau_{\mu}{}^\lam\,\tau_{\lam\nu}+\frac{1}{12}\tau\,\tau_{\mu\nu}+\frac{1}{8}\tau^{\alpha\beta}\tau_{\alpha\beta}
\,h_{\mu\nu}-\frac{1}{24}\tau^2\,h_{\mu\nu}\,.$
In the above, $G_4$ constitutes the effective four-dimensional gravitational constant on the brane; this is also defined
in terms of the fundamental gravitational constant $\kappa_5^2$ and the brane tension $\sigma$. The quantity $\pi_{\mu\nu}$
is the well-known quadratic contribution of $\tau_{\mu\nu}$ \cite{SMS} which here, however, trivially vanishes since 
$\tau_{\mu\nu}=0$. Finally, $T^{(eff)}_{\mu\nu}$ can be interpreted as the effective energy-momentum tensor on the brane.
Together with $E_{\mu\nu}$, they constitute the imprint of the dynamics of the bulk fields---gravitational, and possibly gauge
and scalar fields generating the bulk energy-momentum tensor $T^{(B)}_{MN}$---on the brane.  The components of
$T^{(eff)}_{\mu\nu}$ are given by the following relation
\eq$\label{new-brane-ene-comp}
T^{(eff)}_{\mu\nu}=\frac{1}{\kappa_5^2k}\left[3k^2h_{\mu\nu}-{\Lambda}\,h_{\mu\nu}+\frac{M}{r^3}\left(
\begin{array}{cccc}
 -h_{tt} & 0 & 0 & 0 \\
 0 & - h_{rr} & 0 & 0 \\
 0 & 0 & h_{\theta\theta} & 0 \\
 0 & 0 & 0 & h_{\phi\phi} \\
\end{array}
\right)\right]\,,$
while the components of the tensor $E_{\mu\nu}$, defined in \eqref{E-weyl}, are evaluated to be
\eq$\label{E-weyl-comp}
E_{\mu}{}_{\nu}\Big|_{y\ra 0}=\left(-\frac{Q^2}{r^4}+\frac{M}{r^3}\right)\left(
\begin{array}{cccc}
  -h_{tt} & 0 & 0 & 0  \\
 0 & -h_{rr} & 0 & 0  \\
 0 & 0 & h_{\theta\theta} & 0  \\
 0 & 0 & 0 & h_{\phi\phi} 
\end{array}
\right)\,.$
We notice that $E_{\mu\nu}$ is evaluated infinitesimally close to the brane but not exactly on it, its source
being the five-dimensional Weyl tensor.  Substituting the above relations in \eqref{grav-eqs-br-new}, we obtain
\eq$\label{4dEin}
\bar{G}_{\mu\nu}=\left(\begin{array}{cccc}
  \left(\frac{Q^2}{r^4}+\Lambda_4\right)f(r) & 0 & 0 & 0  \\
 0 & -\left(\frac{Q^2}{r^4}+\Lambda_4\right)\frac{1}{f(r)} & 0 & 0  \\
 0 & 0 & \left(\frac{Q^2}{r^4}-\Lambda_4\right)r^2 & 0  \\
 0 & 0 & 0 & \left(\frac{Q^2}{r^4}-\Lambda_4\right)r^2\sin^2\theta
\end{array}
\right)\,,$
with
\eq$
f(r)\equiv 1-\frac{2M}{r}+\frac{Q^2}{r^2}-\frac{\Lambda_4}{3}r^2\,.$
One can verify that the expression of the Einstein tensor in \eqref{4dEin} matches exactly the Einstein
tensor of the four-dimensional Reissner-Nordstr\"{o}m-(A)dS metric, with $\Lambda_4=\Lambda$ being
the effective cosmological constant on the brane and $Q^2/r^4$ the equivalent of the energy-momentum
tensor component of an electromagnetic field.  Although we have called our five-dimensional black-hole
solution a Reissner-Nordstr\"{o}m-(A)dS one, it is clear that no four-dimensional electromagnetic field has
been---or needed to be---introduced on the brane. The ``charge'' $Q$ is a conserved quantity
carried by the bulk fields and left as an imprint in the four-dimensional spacetime. It is therefore
a tidal charge, as the one accommodated in the black-hole solution of \cite{tidal}, rather than
an ordinary electromagnetic one. It is worth noting that our present analysis can be considered as one
which completes the brane black-hole solution found in \cite{tidal} since it provides the bulk 
description that was lacking from the aforementioned work.

\section{Epilogue}
\label{sec: Ep}

In this work, we have generalised our previous analysis \cite{NK1}, where we studied the localisation of 
a five-dimensional spherically-symmetric, neutral and asymptotically-flat black hole on our brane,
by considering also a cosmological constant and a charge term in the metric function. We have 
preserved the assumption of spherical symmetry in the five-dimensional bulk and by adopting
an appropriate set of spherical coordinates, we have built a black-hole solution with its singularity
strictly residing on the brane. We have performed a careful classification of the horizons that
this background admits, depending on the values of its parameters, and demonstrated that all of 
them have pancake shapes and one after the other get exponentially localised close to the brane.
The bulk gravitational background is everywhere regular, as the calculation of all scalar gravitational
quantities has shown, and reduces to an AdS$_5$ spacetime right outside the black-hole event horizon. 
Our analytically constructed, five-dimensional line-element has all the desired geometric features
of a physical black hole localised close to our brane, and shares the exact same structure as the
line-element employed in \cite{Figueras1} where such a solution was numerically constructed.

In order to support such a geometric background, we need to assume the presence of a bulk
energy-momentum tensor. In terms of the spherical coordinates, this quantity has a minimal structure:
it is diagonal with only two independent components, the energy density $\rho_E$ and tangential pressure
$p_2$, and may be thus described as an anisotropic fluid. Close to and on the brane, both $\rho_E$
and $p_2$ are positive, and respect all energy conditions. However, in order to localise the black-hole
topology close to the brane and prevent the leaking of the singularity along the extra dimension,
a transition needs to take place in the bulk resulting in the violation of both the $\rho_E \geq 0$ and
$\rho_E + p_2 \geq0$ conditions. This violation is only local and takes place within the event horizon
regime in the bulk---soon afterwards, both $\rho_E$ and $p_2$ reduce to constant quantities, which give
rise to the AdS$_5$ spacetime outside the black-hole event horizon. In fact, a general question 
emerges from our analysis as to the nature of the necessary bulk matter that is usually asked to satisfy all energy
conditions at the location of our brane and, at the same time, to support asymptotically -- in the context
of most brane-world models -- an AdS spacetime which by construction violates most energy conditions. 
Are there matter or field configurations that would support and smoothly match these two asymptotic
behaviours?

To this end, we attempted to provide a physical interpretation of the nature of the bulk matter by building a
field-theory model involving scalar and gauge fields living in the bulk. Without determining explicitly
the profiles of these fields---a task that would demand numerical analysis, we obtained the primary
constraints and equations for a viable solution. Although we demonstrated that this scalar-vector model
could indeed reproduce the general structure of the energy-momentum tensor in the bulk, our analysis
also revealed that the gauge, and inevitably the scalar, fields should become phantom-like at the bulk
boundary. The decision on whether a five-dimensional tensor-scalar-vector theory, whose particle
degrees of freedom are well behaved near and on our brane but they turn phantom-like away from it,
is physically acceptable  is still pending. Alternative field theory constructions could also be considered. 
For instance, the negative sign of the energy density of the bulk matter  points perhaps to a non-minimal
gravitational coupling of the fields that takes over at the outskirts of the bulk---the fact that all terms
proportional to the charge $Q$, and therefore sourced by the bulk gauge field, remain always
positive whereas the gravitational terms proportional to $M$ are the ones that cause the energy
density to turn negative seems to agree with this. We plan to study this alternative model in a
follow-up work.

By considering the junction conditions, we have subsequently studied in detail the consistent
embedding of our 3-brane into the bulk geometry we have constructed. We have demonstrated
that no additional matter needs to be introduced on the brane by hand, and that the only energy
content of our brane in the context of the five-dimensional theory is its constant , and positive
self-energy or tension. In fact it is this quantity together with the five-dimensional gravitational
constant that determine the warp parameter of the bulk metric---we note that the warp factor
of the model has the exact same form as the one of the original Randall-Sundrum model, a feature
which also ensures the localisation of gravity close to our brane. These two fundamental quantities
determine also the effective four-dimensional gravitational constant on our brane as the study of
the effective theory on the brane revealed. There, we showed that the combined effect of the
five-dimensional geometry and the bulk matter leaves its imprint on the brane and supports
the Reissner-Nordstr\"{o}m-(A)dS geometry that the four-dimensional observer sees. Let us,
however, stress again that the charge appearing in the metric is a tidal charge, first employed
in the brane construction of \cite{tidal}, rather than an electromagnetic one as it is sourced by the
bulk, gravitational and gauge, fields.  In this sense, our work provides the description of the
bulk geometry which gives rise to the four-dimensional Reissner-Nordstr\"{o}m-type of background
of \cite{tidal}  and which was missing from that analysis.

Apart from the successful localisation of the black-hole geometry close to our brane and the
incorporation of the Randall-Sundrum model in our analysis, our construction supports an
Anti-de Sitter spacetime not only at the bulk boundary but effectively throughout the bulk
regime outside the black-hole event horizon. Therefore, our results could be considered also
in the context of holography \cite{Maldacena,  Gubser2, Witten} and used to study interesting
field-theory phenomena such as chiral symmetry breaking \cite{Pomarol, Alho}, 
confinement/deconfinement \cite{Ballon}, etc. Future directions of work could also address
the stability behaviour of our solution as the Gregory-Laflamme instability arguments \cite{GL}
do not hold here. In previous studies, a stability analysis led also to observable effects such as
echoes of braneworld compact objects \cite{Chakraborty1,Chakraborty2}, as well as other
exotic compact objects \cite{Cardoso,Zachary,Maggio}. A natural question
emerges of whether gravitational waves from black hole mergers or other astrophysical processes
could provide evidence for extra dimensions and distinguish brane-world solutions of this type from the
corresponding four-dimensional ones \cite{Chakraborty3}. The study of the cosmological aspects
of our construction on the brane is also a future direction of research (see, for example,
 \cite{Antonini1, Antonini2}). Also, could we construct alternative
localised black-hole solutions  by considering different forms of the metric function $f(\rho)$,
such as the Schwarzschild-Rindler-(Anti-)de Sitter solution with an additional  linear term  associated
with dark matter or scalar-hair effects \cite{Alestas}, and what would be in that case the profile
of the bulk matter? Is it finally possible to construct rotating brane-world black holes using a similar
process as the one we developed for static brane-world black holes? We plan to return to, at least, 
some of those questions, in future works.

\vspace{1em}
\textbf{Acknowledgements.} The research of T.N. was co-financed by Greece and the European Union 
(European Social Fund- ESF) through the Operational Programme “Human Resources Development, 
Education and Lifelong Learning” in the context of the project “Strengthening Human Resources 
Research Potential via Doctorate Research – 2nd Cycle” (MIS-5000432), implemented by the State 
Scholarships Foundation (IKY).

\appendix

\section{Curvature Invariants}
\label{app: Curv-Inv}
In $(\rho,\chi)$ coordinates, the expressions of the scalar invariants $\mathcal{R}$ and $\mathcal{K}$ are given by
\bal$\label{ricci-2}
\mathcal{R}&=\frac{80}{9} \left(3 k^2-\Lambda\right)^2-\frac{32 k^2 M \left(3 k^2-\Lambda \right) \cos ^2\chi }{\rho }
-\frac{8 \left(27k^3 M \cos^3 \chi +12 k^2-4 \Lambda \right) \left(2 k Q^2 \cos \chi+M\right)}{3 \rho ^3}\nonum\\[0.5mm]
&\hsp+\frac{2 k \cos \chi  \left\{k \cos \chi  \left[3 k^2 \left(9 M^2+16 Q^2\right)-16 \Lambda  Q^2\right]+M \left[9 k^3 M \cos (3 \chi )
+96 k^2-32\Lambda \right]\right\}}{\rho ^2}\nonum\\[0.5mm]
&\hsp+\frac{6 k^2 \cos ^2\chi \left[6 k^2 Q^4 \cos (2 \chi )+6 k^2 Q^4+4 k M Q^2 \cos \chi +17 M^2\right]}{\rho ^4}
+\frac{14 \left(2 k Q^2 \cos \chi +M\right)^2}{\rho ^6}\nonum\\[0.5mm]
&\hsp+\frac{12 k \cos \chi  \left[2 k^2 Q^4 \cos (2 \chi )+2 k^2 Q^4-8 k M Q^2 \cos \chi -5 M^2\right]}{\rho ^5}\,,\\[1.5mm]
\label{riemsq}
\mathcal{K}&=\frac{40}{9} \left(3 k^2-\Lambda\right)^2-\frac{16 k^2 M \left(3 k^2-\Lambda \right) \cos ^2\chi}{\rho }
+\frac{8 k \cos \chi  \left[3 k^2 \left(27 M^2-4 Q^2\right)+4 \Lambda  Q^2\right]}{3\rho^3}\nonum\\[0.5mm]
&\hsp+\frac{4 k \cos \chi \left\{k \cos \chi  \left[3 k^2 \left(9 M^2+4 Q^2\right)-4 \Lambda  Q^2\right]
+M \left[9 k^3 M \cos (3 \chi )+24 k^2-8 \Lambda \right]\right\}}{\rho ^2}\nonum\\[0.5mm]
&\hsp-\frac{4 M \left\{27 k^3 \left[k Q^2 \left(4 \cos (2 \chi )+\cos (4 \chi )+3\right)-2 M \cos (3 \chi )\right]+12 k^2
-4 \Lambda \right\}}{3 \rho ^3}\nonum\\[0.5mm]
&\hsp+\frac{24 k^2 \cos ^2\chi \left[3 k^2 Q^4 \cos (2 \chi )+3 k^2 Q^4-28 k M Q^2 \cos \chi +19 M^2\right]}{\rho ^4}\nonum\\[1mm]
&\hsp+\frac{48 k \cos \chi \left[4 k^2 Q^4 \cos (2 \chi )+4 k^2 Q^4-19 k M Q^2 \cos \chi +5 M^2\right]}{\rho ^5}
+\frac{72Q^4}{\rho^8}\nonum\\[0.5mm]
&\hsp+\frac{8 \left[31 k^2 Q^4 \cos (2 \chi )+31 k^2 Q^4-64 k M Q^2 \cos \chi +11 M^2\right]}{\rho ^6}
-\frac{144 Q^2 \left(M-2 k Q^2 \cos \chi\right)}{\rho ^7}\,,$
while, in $(r,y)$ coordinates, the above expressions take the form
\bal$\label{ricci-2-ry}
\mathcal{R}&=\frac{80}{9} \left(3 k^2-\Lambda\right)^2+\frac{2 k^6 M^2 \left(160-384 e^{k \left| y\right| }
+375 e^{2 k \left| y\right| }-180 e^{3 k \left| y\right| }+36 e^{4 k \left| y\right| }\right)}{\left[k^2r^2
+\left(e^{k|y|}-1\right)^2\right]^3}\nonum\\[2mm]
&\hsp+\frac{32k^4 Q^2 \left(3 k^2
-\Lambda \right) \left(e^{k \left| y\right| }-1\right) \left(3 e^{k \left| y\right| }-5\right)}{3 \left[k^2r^2
+\left(e^{k|y|}-1\right)^2\right]^2}+\frac{8k^8 Q^4 \left(e^{k \left| y\right| }-1\right)^2 \left(10-12 e^{k \left| y\right| }
+9 e^{2 k \left| y\right| }\right)}{\left[k^2r^2+\left(e^{k|y|}-1\right)^2\right]^4}\nonum\\[2mm]
&\hsp-\frac{8k^3M}{3\left[k^2r^2+\left(e^{k|y|}-1\right)^2\right]^{7/2}}\bigg\{40 \Big[3 k^6 r^4
+k^4 \left(3 Q^2-\Lambda  r^4+6 r^2\right)+k^2 \left(3-2 \Lambda  r^2\right)-\Lambda \Big]\nonum\\[2mm]
&\hsp-16 e^{k \left| y\right| } \Big[9 k^6 r^4+k^4 \left(21 Q^2-3 \Lambda  r^4+48 r^2\right)+k^2 \left(39
-16 \Lambda  r^2\right)-13 \Lambda \Big]\nonum\\[2mm]
&\hsp+e^{2 k \left| y\right| } \Big[36 k^6 r^4+k^4 \left(387 Q^2-12 \Lambda  r^4+888 r^2\right)
-148 k^2 \left(2\Lambda  r^2-9\right)-444 \Lambda \Big]\nonum\\[2mm]
&\hsp+e^{3 k \left| y\right| } \Big[-9 k^4 \left(25 Q^2+48 r^2\right)+48 k^2 \left(3 \Lambda  r^2-31\right)
+496 \Lambda \Big]\nonum\\[2mm]
&\hsp+2 e^{4 k \left| y\right| } \Big[9 k^4 \left(3 Q^2+4 r^2\right)-12 k^2 \left(\Lambda  r^2-38\right)
-152 \Lambda \Big]+12 \left(3 k^2-\Lambda \right) e^{5 k \left| y\right| }\left( e^{k \left| y\right| }-8\right)\bigg\}
\,,$
\bal$\label{riemsq-ry}
\mathcal{K}&=-\frac{16k^3 M \left(3 k^2-\Lambda \right) \left(10-12 e^{k \left| y\right| }
+3 e^{2 k \left| y\right| }\right)}{3 \left[k^2r^2+\left(e^{k|y|}-1\right)^2\right]^{3/2}}
-\frac{16k^7 M Q^2 \left(10-28 e^{k \left| y\right| }+39 e^{2 k \left| y\right| }
-30 e^{3 k \left| y\right| }+18 e^{4 k \left| y\right| }\right)}{\left[k^2r^2+\left(e^{k|y|}-1\right)^2\right]^{7/2}}\nonum\\[2mm]
&\hsp+\frac{8 k^6 M^2 \left(20-48 e^{k \left| y\right| }+57 e^{2 k \left| y\right| }
-36 e^{3 k \left| y\right| }+18 e^{4 k \left| y\right| }\right)}{\left[k^2r^2+\left(e^{k|y|}-1\right)^2\right]^{3}}
+\frac{16k^4 Q^2 \left(3 k^2-\Lambda \right) \left(e^{k \left| y\right| }-1\right) \left(3 e^{k \left| y\right| }
-5\right)}{3\left[k^2r^2+\left(e^{k|y|}-1\right)^2\right]^{2}}\nonum\\[2mm]
&\hsp+\frac{8 k^8 Q^4 \left(5-16 e^{k \left| y\right| }+26 e^{2 k \left| y\right| }-24 e^{3 k \left| y\right| }
+18 e^{4 k \left| y\right| }\right)}{\left[k^2r^2+\left(e^{k|y|}-1\right)^2\right]^{4}}+\frac{40}{9} \left(3 k^2-\Lambda\right)^2\,.$

\section{How to remedy the cosmological horizon singularity}
\label{app: CHCS}

The line-element \eqref{5d-metr} in terms of the radial, null coordinates $(u,v)$ which are defined by
\eq$\label{uv-coords}
\left\{\begin{array}{c}
v=t+\rho_*\\[1mm]
u=t-\rho_*
\end{array}\right\}\,,$
takes the form
\eq$\label{uv-metr}
ds^2=\frac{1}{(1+k\rho\cos\chi)^2}\left[-f(\rho)\,dudv+\rho^2d\Omega_3^2\right]\,.$
In the above, the variable $\rho_*$ is determined by the following relation
\eq$\label{rho-star}
\rho_*=\int\frac{d\rho}{f(\rho)}=-\frac{1}{2\kappa_{\ssst{C}}}\ln\bigg|\frac{\rho}{\rho_{\ssst{C}}}-1\bigg|+\frac{1}{2
\kappa_{\ssst{+}}}\ln\bigg|\frac{\rho}{\rho_{\ssst{+}}}-1\bigg|-\frac{1}{2\kappa_{\ssst{-}}}\ln\bigg|\frac{\rho}{\rho_{\ssst{-}}}
-1\bigg|+\frac{1}{2\kappa_4}\ln\bigg|\frac{\rho}{\rho_4}-1\bigg|\,,$ where the integration constant has been set to zero. 
The constants $\rho_{\ssst{C}}$, $\rho_{\ssst{+}}$, $\rho_{\ssst{-}}$, $\rho_4$ are the roots\footnote[9]{It is implied that 
$\rho_1=\rho_{\ssst{C}}$, $\rho_2=\rho_{\ssst{+}}$, $\rho_3=\rho_{\ssst{-}}$.} of the quartic polynomial
$f(\rho)=0$ which for $\Lambda>0$ satisfy the inequality $\rho_{\ssst{C}}>\rho_{\ssst{+}}>\rho_{\ssst{-}}>\rho_4$, 
with $\rho_4<0$. The parameters $\kappa_i$ denote the surface gravity at the corresponding $i_{th}$ horizon located 
at $\rho=\rho_i$ (for more details see \cite{Chambers}). Using the aforementioned roots, the function $f(\rho)$ given
by \eqref{ansatz} can be factorised as follows
\eq$\label{f-rho-fact}
f(\rho)=-\frac{\Lambda}{3}\frac{(\rho-\rho_{\ssst{C}})(\rho-\rho_{\ssst{+}})(\rho-\rho_{\ssst{-}})(\rho-\rho_4)}{\rho^2}\,.$
Combining Eqs. \eqref{rho-star} and \eqref{f-rho-fact}, the function $f(\rho)$ near the cosmological horizon reduces
to 
\eq$\label{f-rho-lim}
\lim_{\rho\ra\rho_{\ssst{C}}^\pm}f(\rho)=\mp2\rho_{\ssst{C}}\kappa_{\ssst{C}}e^{-2\kappa_{C}\rho_*}\,,$
where the minus or plus sign on the right-hand-side depends on the direction from which we approach the cosmological horizon, while
\eq$\label{kappa-c}
\kappa_{\ssst{C}}=\frac{\Lambda}{6}\frac{(\rho_{\ssst{C}}-\rho_{\ssst{+}})(\rho_{\ssst{C}}-\rho_{\ssst{-}})
(\rho_{\ssst{C}}-\rho_4)}{\rho_{\ssst{C}}^2}\,.$
The future cosmological horizon $\rho_{\ssst{C}}$ lies at $t\ra+\infty$ and $\rho_*\ra+\infty$, i.e. at $v\ra+\infty$.
Consequently, by defining the coordinates
\eq$\label{UV-coords}
\left\{\begin{array}{l}
V=-e^{-\kappa_{\ssst{C}}v}\\[1mm]
U=e^{\kappa_{\ssst{C}}u}
\end{array}\right\}\,,$
we can readily see that $V\ra0$ as $v\ra+\infty$. Therefore, using the limit \eqref{f-rho-lim} and the $(U,V)$ coordinates, the
line-element \eqref{uv-metr} near the future cosmological horizon takes the form
\eq$\label{UV-metr}
ds^2\simeq\frac{1}{(1+k\rho\cos\chi)^2}\left[\frac{2\rho_{\ssst{C}}}{\kappa_{\ssst{C}}}\,dUdV+\rho^2d\Omega_3^2\right].$
It is easy to see now that in the above coordinate system the geometry close to the cosmological horizon is completely regular.

\section{Bulk energy-momentum tensor components transformed}
\label{app: En-Mom}

In this section, we will derive the components of the energy-momentum tensor as we change from the set of coordinates
$x^M=\{t,\rho,\chi,\theta,\phi\}$ to the set $x'^M=\{t,r,\theta,\phi,y\}$. We will denote all new quantities with a prime
in order to distinguish them from those in the old coordinates. Thus, using Eq. \eqref{en-mom} we have
\eq$\label{en-mom-ry}
T'^{(B)MN}=(\rho_E+p_2)U'^M\, U'^N+(p_1-p_2)X'^M\, X'^N+p_2\, g'^{MN}\,.$
The quantities $\rho_E$, $p_1$ and $p_2$ are scalars and thus they do not change under a coordinate transformation. Their
expressions  in the new coordinates can be easily obtained from Eqs. \eqref{rho-p1} and \eqref{p2} by using the relations
of Eq. \eqref{new-coords-inv}. However, the vectors $U'^M$ and $X'^M$, defined in  \eqref{U-vec} and\eqref{X-vec},
under the coordinate transformation are transformed as follows
\eq$\label{U-ry}
U'^M=\frac{dx'^M}{dx^A}\,U^A=\frac{dx'^M}{dt}\,U^t=\frac{e^{k|y|}}{\sqrt{f(r,y)}}\,\del^M{}_t\,,$
\eq$\label{X-ry}
X'^M=\frac{dx'^M}{dx^A}\,X^A=\frac{dx'^M}{d\rho}\,X^\rho=
\left[r^2+\frac{\left(e^{k|y|}-1\right)^2}{k^2}\right]^{-1/2}e^{k|y|}\sqrt{f(r,y)}\left(r\,\del^{M}{}_r+\frac{1-e^{-k|y|}}{k}\del^{M
}{}_y\right)\,.$
In the above, the function $f(r,y)$ is given in Eq. \eqref{f-ry}. One can also verify that $U'^MU'^Ng'_{MN}=-1$ and
$X'^MX'^Ng'_{MN}=1$,  where $g'_{MN}$ is evaluated from the line-element \eqref{metr-r-y}. Then, for the mixed components
of the energy-momentum tensor $T'^{(B)M}{}_N$, we obtain
\eq$\label{en-mom-mix-ry}
T'^{(B)M}{}_N=T'^{(B)ML}g'_{LN}=(\rho_E+p_2)U'^M\, U'^t\,g'_{tN}+(p_1-p_2)X'^M\left( X'^r\,g'_{rN}+X'^y\,g'_{yN}\right)+p_2\, 
\del^{M}{}_{N}\,.$
Using Eqs. \eqref{U-ry} and \eqref{X-ry} in Eq. \eqref{en-mom-mix-ry}, it is straightforward to calculate the non-zero mixed
components of the energy-momentum tensor in the new coordinate system. These read:
\bal$
T^{(B)t}{}_t&=-\rho_E(r,y)=\frac{1}{\kappa_5^2}\left\{2(3k^2-\Lambda)-\frac{3Mk^3\left(4-3e^{k|y|}\right)}{\left[k^2r^2+\left(e^{k|y|}-1\right)^2\right]^{3/2}}-\frac{6Q^2k^4\left(e^{k|y|}-1\right)}{\left[k^2r^2+\left(e^{k|y|}-1\right)^2\right]^{2}} \right\}\,,$
\bal$T^{(B)r}{}_r&=\frac{1}{\kappa_5^2}\left\{2(3k^2-\Lambda)+\frac{3k^3M \left\{-4\left(1+k^2r^2\right)+e^{k|y|}\left[14-2e^{k|y|}
\left(9-5e^{k|y|}+e^{2k|y|}\right)+3k^2r^2\right]\right\}}{\left[k^2r^2+\left(e^{k|y|}-1\right)^2\right]^{5/2}}\right.\nonum\\[2mm]
&\hspace{2.73cm}\left.+\frac{6 k^4Q^2 \left(e^{k \left| y\right| }-1\right) \left(3 e^{k \left| y\right| }-3 e^{2 k \left| y\right| }
+e^{3 k \left| y\right| }-k^2 r^2-1\right)}{\left[k^2r^2+\left(e^{k|y|}-1\right)^2\right]^{3}}\right\}\,,\\[5mm]
T^{(B)r}{}_y&=e^{2k|y|}\,T^{(B)y}{}_r=\frac{3k^4 r\, e^{2k|y|}}{\kappa_5^2}\left\{ \frac{M \left(e^{k \left| y\right| }-1\right) 
\left(2 e^{k \left| y\right| }-3\right)}{\left[k^2r^2+\left(e^{k|y|}-1\right)^2\right]^{5/2}}-\frac{2 k Q^2 \left(e^{k \left| y\right| }
-1\right)^2}{ \left[k^2r^2+\left(e^{k|y|}-1\right)^2\right]^{3}}\right\}\,,\\[5mm]
T^{(B)\theta}{}_\theta&=T^{(B)\phi}{}_\phi=p_2(r,y)=\frac{1}{\kappa_5^2}\left\{2(3k^2-\Lambda)+\frac{6Mk^3\left(e^{k|y|}-1\right)
\left(2-e^{k|y|}\right)}{\left[k^2r^2+\left(e^{k|y|}-1\right)^2\right]^{3/2}}+\frac{6Q^2k^4\left(e^{k|y|}-1\right)^2}{\left[k^2r^2
+\left(e^{k|y|}-1\right)^2\right]^{2}} \right\}\,,\\[5mm]
T^{(B)y}{}_y&=\frac{1}{\kappa_5^2}\left\{2(3k^2-\Lambda)+ \frac{3k^3M\left(e^{k \left| y\right| }-1\right) \left[e^{k \left| y\right| } 
\left(3 e^{k \left| y\right| }-2 k^2 r^2-7\right)+4\left(1+ k^2   r^2\right)\right]}{\left[k^2r^2+\left(e^{k|y|}-1\right)^2\right]^{5/2}}\right.\nonum\\[2mm]
&\hspace{2.73cm}\left.+\frac{6 k^4 Q^2 \left(e^{k \left| y\right| }-1\right)^2\left(1+k^2r^2-e^{k \left| y\right| }\right)}{\left[k^2r^2+\left(e^{k|y|}-1\right)^2\right]^{3}}\right\}\,.$

\section{Brane energy-momentum tensor in terms of the extrinsic curvature}
\label{app: Br-Ene}
From Eq. \eqref{jc2}, we have
\gat$ \label{app-br-ene-1}
h^{\mu\nu}[K_{\mu\nu}]=
[h^{\mu\nu}K_{\mu\nu}]=-\kappa_5^2 \left(h^{\mu\nu}T^{(br)}_{\mu\nu}-\frac{1}{3}\,h^{\mu\nu}h_{\mu\nu}\,T^{(br)}\right)\Ra
T^{(br)}=\frac{3}{\kappa_5^2}\,[K]\,.$
Using then Eq. \eqref{app-br-ene-1} in Eq. \eqref{jc2}, we obtain
\gat$ \label{app-br-ene-2}
[K_{\mu\nu}]=-\kappa_5^2\left(T^{(br)}_{\mu\nu}-\frac{1}{3}h_{\mu\nu}\,\frac{3}{\kappa_5^2}[K]\right)\Ra
T^{(br)}_{\mu\nu}=-\frac{1}{\kappa_5^2}\left([K_{\mu\nu}]-h_{\mu\nu}[K]\right)\,.$

\addcontentsline{toc}{section}{References}
\bibliography{Bibliography}{}
\bibliographystyle{utphys}

\end{document}